\documentclass[9pt]{article}
\usepackage[a4paper,total={7in, 10in}]{geometry}
\usepackage{authblk}
\usepackage{dsfont}
\usepackage{mathtools}
\usepackage{amsmath}
\usepackage{amssymb}
\usepackage{mymacros}
\usepackage[section]{placeins} \usepackage{lineno}
\title{
	A differentiable forward model for the concurrent, multi-peak Bragg coherent x-ray diffraction imaging problem
	}

\author[1]{S. Maddali}
\author[1]{T. D. Frazer}
\author[1,2]{N. Delegan}
\author[1]{K. J. Harmon}
\author[1,2]{S. E. Sullivan}
\author[3]{M. Allain}
\author[4]{W. Cha}
\author[5]{A. Dibos}
\author[1]{I. Poudyal}
\author[4]{S. Kandel}
\author[6,$\dagger$]{Y. S. G. Nashed}
\author[1,2,7]{F. J. Heremans}
\author[1]{H. You}
\author[1,7]{Y. Cao}
\author[1,*]{S. O. Hruszkewycz}

\affil[1]{Materials Science Division, Argonne National Laboratory Lemont, IL 60439 (USA)}
\affil[2]{Center for Molecular Engineering, Argonne National Laboratory, Lemont, IL 60439 (USA)}
\affil[3]{Aix-Marseille Univ, CNRS, Centrale Marseille, Institut Fresnel, Marseille, France}
\affil[4]{X-ray Science Division, Argonne National Laboratory, Lemont, IL 60439 (USA)}
\affil[5]{Center for Nanoscale Materials, Argonne National Laboratory, Lemont, IL 60439 (USA)}
\affil[6]{Mathematics \& Computer Science Division, Argonne National Laboratory, Lemont, IL 60439 (USA)}
\affil[7]{Pritzker School of Molecular Engineering, University of Chicago, 5640 South Ellis Avenue, Chicago, IL 60637 (USA)}
\affil[$\dagger$]{Currently at SLAC National Accelerator Laboratory, CA (USA)}

\affil[*]{Corresponding author: shrus@anl.gov}

\date{}

\begin{document}

\maketitle
\begin{abstract}
	We present a general analytic approach to spatially resolve the nano-scale lattice distortion field of strained and defected compact crystals with  Bragg coherent x-ray diffraction imaging (BCDI).
Our approach relies on fitting a differentiable forward model simultaneously to multiple BCDI datasets corresponding to independent Bragg reflections from the same single crystal. 
It is designed to be faithful to heterogeneities that potentially manifest as phase discontinuities in the coherently diffracted wave, such as lattice dislocations in an imperfect crystal. 
We retain fidelity to such small features in the reconstruction process through a Fourier transform -based resampling algorithm designed to largely avoid the point spread tendencies of commonly employed interpolation methods. 
The reconstruction model defined in this manner brings  BCDI reconstruction into the scope of explicit optimization driven by automatic differentiation. With results from simulations and experimental diffraction data, we demonstrate significant improvement in the final image quality compared to conventional phase retrieval, enabled by explicitly coupling multiple BCDI datasets into the reconstruction loss function.  \end{abstract}

\section{Introduction}											\label{S:intro}		Bragg coherent diffraction imaging (BCDI) is a synchrotron-based characterization technique that utilizes coherent x-ray illumination and 3D phase retrieval algorithms to interrogate structural heterogeneities in crystalline materials on the scale of tens of nanometers~\cite{Vartanyants1997,Robinson2001,Miao2015}. 
The Bragg diffraction geometry is sensitive to the distortion field of the atomic lattice, components of which are encoded in the complex-valued phase of the scattered wave.
Conventional BCDI seeks to spatially resolve individual projections of the lattice distortion field by applying phase retrieval methods to a single three-dimensional Bragg diffraction pattern. 
This diffraction pattern corresponds to interfering x-rays reflected from a single family of atomic planes within a crystal. 
The corresponding component of the elastic strain tensor may be subsequently obtained by spatial differentiation of the retrieved phase.
This technique has greatly benefited research that non-destructively interrogates crystalline materials in which properties and behavior are influenced by lattice strain, such as charging cycle -induced degradation of solid-state batteries~\cite{Ulvestad2015,Singer2018}, catalytic reactions at metal nanoparticle interfaces~\cite{Kawaguchi2019} and quantum sensors~\cite{Kraus2014,Hruszkewycz2018}.  
Numerically, phase retrieval typically involves fixed point iterative projections (FPIPs) acting on an initial guess object.
The solution is  alternately constrained to an assumed support mask in real space and the measured signal in Fourier space~\cite{Gerchberg1972,Fienup1982}.
The initial support estimate is intermittently refined using a `shrinkwrap' algorithm~\cite{Marchesini2003}. 
The iterations proceed until the solution is deemed converged. 
A given combination of FPIP algorithms for a particular reconstruction task, or `recipe', is determined by trial-and-error and is typically validated only in hindsight upon successful reconstruction. 
This method of recipe design becomes increasingly difficult when reconstructing defected crystals, for which convergence is far less stable, and sometimes even elusive. 

Recent efforts aim to directly retrieve the 3D lattice distortion field by concurrently solving the phase retrieval problems for multiple independent BCDI datasets and then fitting the lattice distortion to the reconstructed phase profiles. 
Proof-of-concept demonstrations so far have used combinations or variants of established FPIPs in their reconstruction algorithms~\cite{Newton2020,Gao2021,Hofmann2020,Wilkin2021}. 
The majority of these are purely simulation-based with highly idealized scattering crystals and distortion fields. 
This variety of multi-reflection BCDI (MR-BCDI) has only recently been demonstrated with experimental data~\cite{Wilkin2021}. 

In this paper we present an gradient descent -based solution for the general MR-BCDI problem which offers the following major advantages over current practices: 
\begin{enumerate}
	\item \textbf{Feature and data fidelity}:
		In our reconstructions we maintain fidelity to small features within the scattering crystals ($\sim$ 1 pixel), as well as to the noise statistics in the diffraction data itself, by implementing a Fourier transform -based re-gridding method in real space, and avoiding data re-gridding altogether. 
		Grid interpolation either in real or Fourier space is typically necessary in BCDI to reconcile signal digitization with a chosen reconstruction grid. 
		Current methods that employ straightforward multi-linear or polynomial interpolation are susceptible to point spread artifacts, which our method greatly minimizes. 
		The significance of our Fourier-based approach is the ability to reconstruct crystals with higher defect and dislocation densities than the usual FPIP approach.
	\item \textbf{Flexible convergence}:
		Stochastic gradient descent implemented over the whole body of BCDI data affords us a flexible path to convergence to the desired solution, in the manner of training of present-day machine learning models~\cite{Saad1998,Duchi2011,Kingma2014}. 
		In contrast, FPIP approaches using trial-and-error recipes commit to an arbitrary rigid progression in solution space~\cite{Marchesini2007}, punctuated by uncontrolled `resets' due to shrinkwrap. 
		The stochastic optimization approach is shown to be far more reliable for convergence than a typical trial-and-error recipe. 
	\item \textbf{Reconstruction quality}:
		We demonstrate an improvement in the reconstruction quality enabled by explicit coupling of the BCDI data sets into the loss function.
		This forward model approach is fundamentally different from intermittent or \emph{post hoc} bootstrapping of the concurrently calculated phases through the underlying lattice distortion~\cite{Newton2020, Wilkin2021}.
\end{enumerate}

This paper is organized as follows.  
Section~\ref{S:multiref} describes the MR-BCDI loss function over which the optimization is performed, similar to the well-known Gaussian loss function used in phase retrieval~\cite{Godard2012}. 
It is formulated in terms of the lattice distortion field suitably digitized to be consistent with each of the BCDI signals. Section~\ref{S:results} demonstrates reconstruction results from simulated as well as experimental data, the latter obtained from a specially prepared sub-micron sized silicon carbide crystal for quantum information and sensing applications. 
Section~\ref{S:results} also compares the results to conventional phase retrieval reconstructions of the same data. 
Section~\ref{S:discuss} discusses the place of our algorithm in the current landscape of coherent diffraction imaging techniques and its impact on future BCDI experiments, along with an assortment of issues like stability, support update, interpretability and optimization constraints. 
Finally, the Methods section~\ref{S:ftresmpl} details the various Fourier transform-based operations employed in the multi-reflection forward model, as well as specifics of the silicon carbide sample preparation. 

 \section{The multi-reflection forward model}					\label{S:multiref}	
\begin{figure}[ht!]
	\centering
	\includegraphics[width=0.75\textwidth]{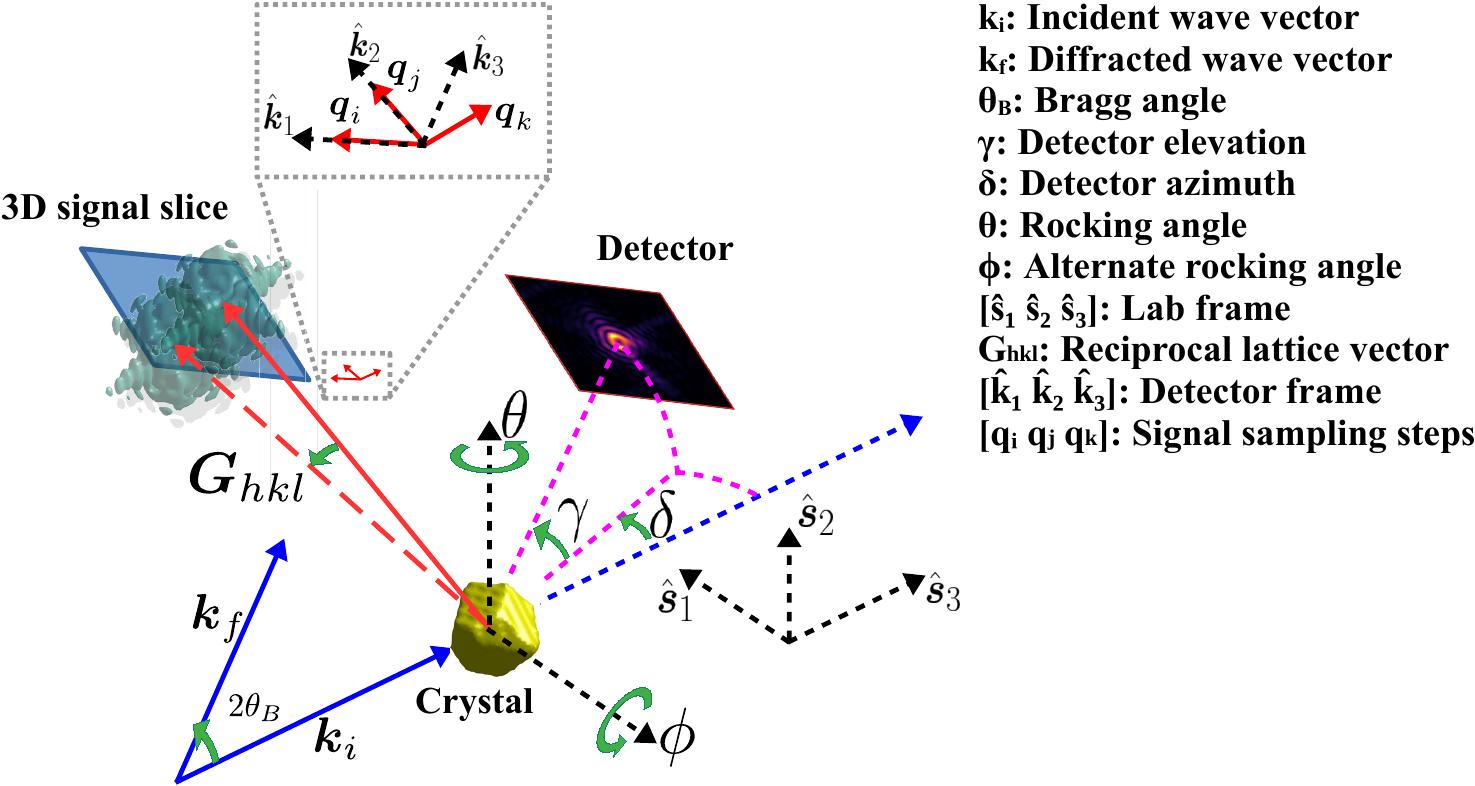}
	\caption{
		BCDI setup at Beamline 34-ID-C of the Advanced Photon Source showing the laboratory and detector frames $[\unitvector{s}_1~\unitvector{s}_2~\unitvector{s}_3]$ and $[\unitvector{k}_1~\unitvector{k}_2~\unitvector{k}_3]$ for a single Bragg condition. 
	}
	\label{fig.schematic}
\end{figure}

To describe the the distortion field we wish to reconstruct, we nominally adopt the right-handed ortho-normal laboratory frame $[\unitvector{s}_1~\unitvector{s}_2~\unitvector{s}_3]$ commonly used at the Advanced Photon Source (Figure~\ref{fig.schematic}). 
The following formalism can be extended generally to lab frames at other synchrotron facilities. 
Here, $\unitvector{s}_3$ is a unit vector pointing downstream along the incident beam and $\unitvector{s}_2$ is vertically upward. 
In this frame, we define the following quantities: 
\begin{itemize} 
	\item	
		The scalar field $\mathcal{A}:\mathbb{R}^3 \rightarrow \mathbb{R}$ and the 3D vector field $\bs{u}:\mathbb{R}^3 \rightarrow \mathbb{R}^3$, respectively denoting the local electron density 
and the vector displacements of the atoms from their equilibrium positions, at location $\bs{x} \in \mathbb{R}^3$. 
		For optimization purposes we assume $\mathcal{A}(\bs{x}) \in [0,1]$ in arbitrary units.
$\mathcal{A}(\bs{x}) = 0$ and $\bs{u}(\bs{x}) = \bs{0}$ is enforced outside a user-defined bounding box $\mathcal{V}$. 
$\mathcal{A}$ and $\bs{u}$ are discretized on a user-defined grid within 
$\mathcal{V}$, given by a voxel size $s_0$ along each of the Cartesian axes $[\unitvector{s}_1~\unitvector{s}_2~\unitvector{s}_3]$. 
\item	
		The reciprocal vectors of the crystal lattice corresponding to the BCDI data sets ($M$ in number) are denoted by the set $\{\bs{G}_i\}_{i=1}^M$. 
		The physical units of the $\bs{G}_i$ are chosen to be the inverse of those chosen for $\bs{u}(\bs{x})$, \emph{i.e.}, $\norm{\bs{G}_i} = 1/d_i$ (crystallographers' convention, as opposed to the physicists' convention of $2\pi/d_i$), where $d_i$ is the corresponding atomic plane spacing.
\item	
		The $3\times 3$ rotation matrices $\{\mathcal{R}_i\}_{i=1}^M$, to bring the crystal lattice into the $\bs{G}_i$ Bragg condition (hereafter referred to as the $i$'th \emph{mounting position}). 
	\item
		The $3\times 3$ matrices $\left\{\bs{B}_\text{real}^{(i)}\right\}_{i=1}^M$ whose columns form a basis of 3D real-space sampling vectors, corresponding to the $i$'th mounting position. 
		These are determined by the diffraction geometry specific to each Bragg reflection~\cite{Maddali2020a}. 
\item	
		Dataset-dependent scaling factors $\left\{\chi_i \left| \chi_i\in \mathbb{R}\right.\right\}_{i=1}^M$ for the electron density $\mathcal{A}$ that controls total energy from each of the $i$ diffracted waves that is incident upon the detector over the course of the scan. 
		We note that for a measured BCDI dataset, this total energy depends not only on the structure factor of the scattering crystal but also the acquisition time for each image in the scan. 
\end{itemize}

These definitions lead to the following deductions: 
\begin{enumerate}
	\item	
The complex-valued object wave consistent with the projection of $\bs{u}(\bs{x})$ along $\bs{G}_i$, 
\emph{prior to rotating into the Bragg condition}, is simply: $\psi^{(i)}(\bs{x}) = \mathcal{A}(\bs{x})\exp[\iota 2\pi \bs{G}_i^T\bs{u}(\bs{x})]$.
		After the crystal is mounted into the $i$'th position, the \emph{rotated} version of this complex-valued object, \emph{i.e.}, $\psi^{(i)}(\mathcal{R}_i^{-1}\bs{x}) \equiv \mathcal{A}(\mathcal{R}_i^{-1}\bs{x})\exp[\iota 2\pi \bs{G}_i^T\bs{u}(\mathcal{R}_i^{-1}\bs{x})]$, represents the 3D object wave to be eventually propagated to the far field. 
		We note that the phase of $\psi^{(i)}(\mathcal{R}_i^{-1}\bs{x})$ already accounts for the rotation $\mathcal{R}_i$ being applied to the crystal lattice and therefore to $\bs{G}_i$, in the following manner: 
		\begin{align*}
			\bs{G}_i &\xrightarrow[]{\text{rotated by }\mathcal{R}_i} \mathcal{R}_i \bs{G}_i \tag{Transformation of single vector} \\
			\bs{u}(\bs{x}) &\xrightarrow[]{\text{rotated by }\mathcal{R}_i} \mathcal{R}_i \bs{u}\left(\mathcal{R}_i^{-1}\bs{x}\right) \tag{Transformation of vector field} \\
			\therefore \angle\left[\psi^{(i)}\left(\mathcal{R}_i^{-1}\bs{x}\right)\right] &= 2\pi\left(\bs{G}_i^T\mathcal{R}_i^T\right)\left[\mathcal{R}_i\bs{u}\left(\mathcal{R}_i^{-1}\bs{x}\right)\right] \\
			&= 2\pi\bs{G}_i^T\bs{u}\left(\mathcal{R}_i^{-1}\bs{x}\right) \tag{since $\mathcal{R}_i^T\mathcal{R}_i = \mathds{1}$, the identity matrix}
		\end{align*}
		This last expression denotes the rotated phase field within the bounding box $\mathcal{V}$.
\item
		With the definition of $\bs{B}_{\text{real}}^{(i)}$ above, the $i$'th \emph{digital} object wave may be indexed by an integer column vector $\bs{m} \in \mathbb{Z}^3$:
		\begin{align}
			\psi_{\bs{m}}^{(i)} 	= \psi^{(i)}\left(\mathcal{R}_i^{-1}\bs{B}_\text{real}^{(i)}\bs{m}\right)
									= A\left(\mathcal{R}_i^{-1}\bs{B}_\text{real}^{(i)}\bs{m}\right) \exp\left[
										\iota 2\pi \bs{G}_i^T\bs{u}\left(\mathcal{R}_i^{-1}\bs{B}_\text{real}^{(i)}\bs{m}\right)
										\right]	\label{eq.rotate_resampled}
		\end{align}
Assuming negligible cyclic aliasing, the discrete Fourier transform of $\psi_{\bs{m}}^{(i)}$ from Eq.~\eqref{eq.rotate_resampled} is guaranteed to return the continuous Fourier transform $\Psi^{(i)}$ of $\psi^{(i)}(\mathcal{R}^{-1}_i\bs{x})$, sampled on the conjugate grid $\bs{B}^{(i)}_{\text{recip}} \bs{n}$, where $\bs{n} \in \mathbb{Z}^3$. 
\begin{equation}\label{eq.conjugatetransform}
			\psi^{(i)}_{\bs{m}} = \psi^{(i)}\left(\mathcal{R}_i^{-1}\bs{B}_{\text{real}}^{(i)} \bs{m}\right)
			\ftift
			\Psi^{(i)}\left(\mathcal{R}_i^{-1}\bs{B}_{\text{recip}}^{(i)} \bs{n}\right) = \Psi^{(i)}_{\bs{n}}
		\end{equation}
		This assumption of cyclic aliasing is discussed briefly in Section~\ref{S:discuss}
		For correctly computed $\bs{B}_\text{real}^{(i)}$ (typically by calculating $\bs{B}_\text{recip}^{(i)}$ first, see Ref.~\cite{Maddali2020a}), 
the measured coherent diffraction data in the far field will be correctly represented by $\Psi_{\bs{n}}^{(i)}$. 
		The relation between each pair of $\bs{B}^{(i)}_\text{real}$ and $\bs{B}^{(i)}_\text{recip}$ depends on the specific diffraction geometry and the size of the signal-space array chosen to encompass the signal~\cite{Maddali2020a}.
		This conjugate basis formulation serves to conveniently bring some of the capabilities of existing tools such as \emph{xrayutilities}~\cite{Maddali2020c,Kriegner2013} into the scope of automatic differentiation. 

		In our algorithm we choose to solve for the primary quantities of interest, namely $\mathcal{A}$ and $\bs{u}$, in the laboratory frame, but apply our Fourier-based re-gridding algorithm to $\psi^{(i)}_{\bs{m}}$  in each of the $i$ detector frames in turn (\emph{i.e.}, each of the $M$ coordinate frames $[\unitvector{k}_1~\unitvector{k}_2~\unitvector{k}_3]_i$ from Figure~\ref{fig.schematic}), 
after appropriate transformation to that frame.
		We note that Eq.~\eqref{eq.rotate_resampled} makes no explicit reference to the coordinate frame in which these quantities are represented and therefore affords us this flexibility. 
		In Section~\ref{S:ftresmpl}, we switch to a particular detector frame to describe Fourier-based interpolation.
	\item
		The global scaling factor $\chi_i$ in single-reflection BCDI is trivially a constant determined by the signal strength, but cannot be ignored in the MR-BCDI forward model. 
In our algorithm we treat the $\chi_i$ as additional parameters to be optimized, to mathematically reconcile the different signal energies with the constraint that $\mathcal{A} \in [0,1]$ (since they are related through the Fourier transform). 
\end{enumerate}

The operation defined by $T_i \equiv \mathcal{R}_i^{-1}\bs{B}_\text{real}^{(i)}$ in Eq.~\eqref{eq.rotate_resampled} denotes a re-gridding transformation of the object wave achieved by the combination of (i) rotation into the $i$'th Bragg condition followed by (ii) resampling to achieve the effect of $\bs{B}_\text{real}^{(i)}$, starting from the \emph{discretized} $\mathcal{A}(\bs{x})$ and $\bs{u}(\bs{x})$ fields. 
This is achieved by our Fourier interpolation method described in Section~\ref{S:ftresmpl}. 
With this in mind, the formal MR-BCDI inverse problem is set up in terms of a loss function $\mathcal{L}_\text{multi}$ defined by: \begin{align}
	\mathcal{L}_\text{multi}[A, \bs{u}] 	&=		\sum\limits_{i=1}^M \frac{1}N_\text{vox} \sum\limits_{\bs{n}} \left[
													\norm{\mathcal{F}\left\{\chi_i \mathcal{A}(T_i\bs{m}) e^{\iota 2\pi \bs{G}_i^T\bs{u}(T_i\bs{m})}\right\}}_{\bs{n}} - \left(I_{\bs{n}}^{(i)}\right)^{1/2}
													\right]^2 \label{eq.objfun_multi} \\
	\mathcal{A}^\star, \bs{u}^\star, \bs{\chi}^\star	&=		\arg\min\limits_{\mathcal{A},\bs{u},\bs{\chi}} \mathcal{L}_\text{multi}\left[\mathcal{A}, \bs{u}, \bs{\chi}\right] \label{eq.optimization_multi}
\end{align}
where $\bs{\chi}$ represents the set of $M$ scaling factors. 
$I^{(i)}_{\bs{n}}$ is the measured diffraction data corresponding to the $i$'th Bragg reflection and $\bs{n} \in \mathbb{Z}^3$ ranges over the pixels in each $I^{(i)}_{\bs{n}}$.
Eq.~\ref{eq.objfun_multi} represents the mean error in signal amplitude per BCDI data set, aggregated over the number of data sets. 
Here, $N_\text{vox}$ is the number of voxels in a single BCDI data set ($128^3 = 2097152$ throughout this paper). 
In our stochastic gradient descent scheme, the outermost summation in Eq.~\eqref{eq.objfun_multi} was evaluated either over all the $M$ BCDI data sets, or a subset thereof. 
It is an extension of the familiar phase retrieval loss function typically minimized by FPIP~\cite{Godard2012}, based on the square root transformation of the measured data for robustness.
Further, for Eq.~\eqref{eq.optimization_multi} to be over-determined in $\bs{u}(\bs{x})$, we require $M \geq 3$~\cite{Newton2020,Hofmann2020,Gao2021,Kandel2021a}. 
To analytically constrain $\mathcal{A} \in [0, 1]$ in practice, we define an activation function $\alpha: \mathbb{R}^3 \rightarrow \mathbb{R}$ such that: 
\begin{equation}\label{eq.tanh}
	\mathcal{A}(\bs{x}) \equiv \mathcal{A}[\alpha(\bs{x})] = \frac{1}{2}\left[
		1 + \tanh\left(
		\frac{\alpha(\bs{x})}{\alpha_0}
		\right)
	\right]
\end{equation}
With this definition, we see that $\mathcal{A} \simeq 0$ for $\alpha(\bs{x}) \ll 0$ and $\mathcal{A} \simeq 1$ for $\alpha(\bs{x}) \gg 1$. 
The transition between $0$ and $1$ in the neighborhood of  $\alpha(\bs{x}) = 0$ is controlled by the user-defined positive constant $\alpha_0$.
\begin{figure}[ht!]
	\centering
	\includegraphics[width=0.35\textwidth]{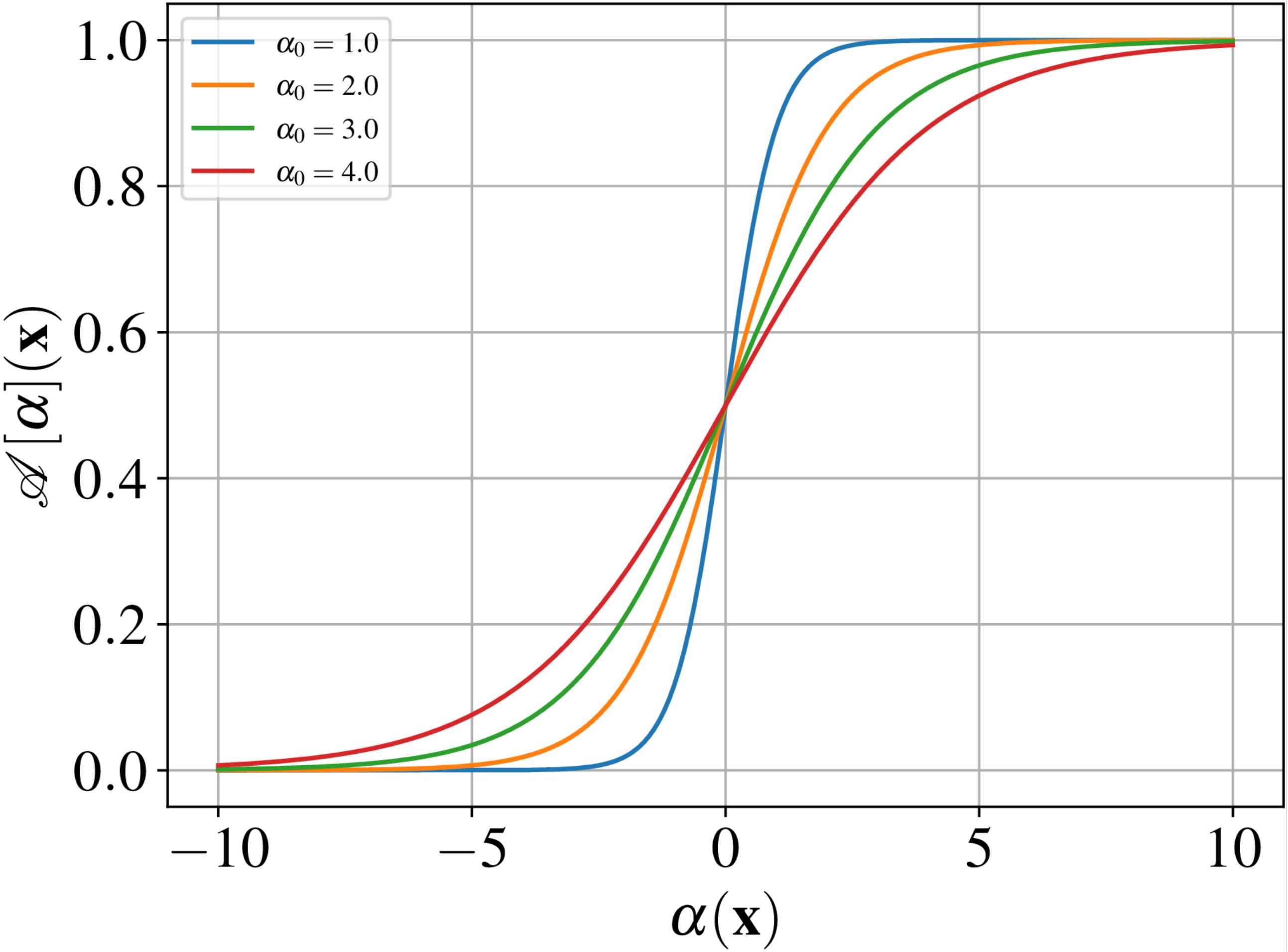}
	\caption{
		Activation function $\alpha(\bs{x})$ used to constrain $\mathcal{A}(\bs{x}) \in [0,1]$, for different activation parameters $\alpha_0$. 
	}
	\label{fig.activation}
\end{figure}
Qualitatively, a larger value of the hyperparameter $\alpha_0$ results in a smaller gradient descent step for $\mathcal{A}$ at $\bs{x}$ (as is seen by differentiating Eq.~\eqref{eq.tanh} with respect to $\alpha$), and a more gradual and controlled approach of $\mathcal{A}$ to its intended value. 
Smaller $\alpha_0$ results in $\mathcal{A}$ converging through sporadic switching between values close to $0$ and $1$, and as such is less controlled. 
$\alpha_0$ is therefore interpreted as parameterizing how strongly the gradient descent solution oscillates around the local minimum in the subspace of $\alpha(\bs{x})$ as it approaches convergence.
We found that $\alpha_0 = 1$ best served our global convergence rate, and have used this for all reconstructions in this paper. 
Given this, the actual optimization problem to be solved in practice is: 
\begin{align}\label{eq.opt_actual}
	\alpha^\star, \bs{u}^\star, \bs{\chi}^\star &= \arg\min\limits_{\alpha, \bs{u}, \bs{\chi}} 
	\mathcal{L}_\text{multi}\left[\mathcal{A}[\alpha], \bs{u}, \bs{\chi}\right] \\
	\mathcal{A}^\star &= \mathcal{A}[\alpha^\star]
\end{align}
We also note that this optimization procedure as described does away with the use of FPIP algorithms altogether. 
Significantly, it also avoids shrinkwrap for support update, provided that the initial bounding box $\mathcal{V}$ is chosen judiciously.

The size of $\mathcal{V}$ relative to the buffered array is an important factor in overdetermining the optimization problem. 
This is seen as follows. 
We assume $\mathcal{V}$ to have an edge size of $n$ voxels, within a buffered array of size $N$ (with $n < N$). 
Therefore we have $M + 4n^3$ unknowns ($\mathcal{A}$, $\bs{u}$ and $\bs{\chi}$ combined), where $M$ is the number of BCDI data sets. 
The number of measurements (photon counts) is given by: $MN^3$. 
The MR-BCDI problem is therefore over-determined by a factor of $MN^3/( M+4n^3)$. In the case of our dislocation-free crystal, $M = 5$, $N = 128$ and $n = 46$ and therefore the overdetermination factor is $\sim 27$ which is significantly higher than the minimum required $1$ and therefore highly desirable. 
Further, since our MR-BCDI reconstruction uses the fast Fourier transform, the issue of signal oversampling must be addressed for a truly well-conditioned inverse problem. 
This is achieved by ensuring that the bounding box $\mathcal{V}$, \emph{after transformation into each of the $M$ detector frames}, spans $\leq N/2$ voxels along each array axis. 
The transformed $\mathcal{V}$ is no longer cubic, but in general sheared into a parallelopiped. 

Lastly, we note that the optimal solution in Eq.~\eqref{eq.optimization_multi} does not preclude the \emph{universal twin} solution, akin to the twin image degeneracy in single-reflection BCDI~\cite{Fienup1986}. 
In other words, the crystal defined by the transformations $\mathcal{A}(\bs{x}) \rightarrow \mathcal{A}(-\bs{x})$ and $\bs{u}(\bs{x}) \rightarrow -\bs{u}(-\bs{x})$ is also a solution to the MR-BCDI problem, in that it results in the twinned version of the object wave for \emph{all} Bragg conditions, and therefore the same inferred diffraction patterns. 
In this sense, MR-BCDI is still susceptible to a reconstruction degeneracy. A related benefit of this coupled reconstruction approach over traditional FPIP methods is that the global optimization will converge to either the physically correct structure or the universal twin structure automatically, without requiring manual intervention to align individual phase retrieval instances.

 \section{Results}												\label{S:results}	In this section we present three reconstruction results with the MR-BCDI technique described above. 
Two of these are with numerically synthesized digital nanocrystals of gold, with coherent diffraction signal simulated in the far field. 
In these numerical studies, one crystal has a slowly varying continuous distortion field (\emph{i.e.}, strained with no line defects), and the other has a discontinuous, winding  displacement field consistent with two orthogonal screw dislocations within the crystal. 
The third reconstruction result is from experimental diffraction data of a silicon carbide (SiC) nanocrystal consisting of six Bragg reflections measured using BCDI at Beamline 34-ID of the Advanced Photon Source. In each case, we present the MR-BCDI reconstruction of the local electron density $\mathcal{A}(\bs{x})$ and the lattice deformation $\bs{u}(\bs{x})$. 
The MR-BCDI reconstruction software was written as a Python module for GPU hardware using the Adam optimizer routine bundled in PyTorch~\cite{Paszke2019} and is available for general use~\cite{Maddali2022}. 
All reconstructions were performed using a single Nvidia Tesla P100 GPU with 16 GB of RAM. 

We have adopted the matrix coordinate convention for all the cross-section plots in this paper. 
As an example, a cross-section image labeled as $X-Y$ implies that the X-axis is directed from top to bottom along the image, and the Y-axis from left to right. The third axis (in this example, the Z-axis) emerges out of the plane of the figure in order to maintain right-handedness. 
The $X$, $Y$ and $Z$ axes correspond to the laboratory frame of reference (\emph{i.e.}, the $\unitvector{s}_1$, $\unitvector{s}_2$ and $\unitvector{s}_3$ directions in Figure~\ref{fig.schematic} respectively). 
With this convention, we ensure that the reference frame of the reconstructed crystal is consistent with the programmed order of the array axes in the multi-dimensional FFT routines in Python and PyTorch.

\subsection{Simulated crystal without dislocations}	\label{SS:nodisloc}	A synthetic crystal with arbitrary facets was generated on a Cartesian grid of size $128 \times 128 \times 128$ voxels, with a voxel size of $s_0 = 12$ nm along each axis in the laboratory frame ($[\unitvector{s}_1~\unitvector{s}_2~\unitvector{s}_3]$ in Figure~\ref{fig.schematic}). 
A face-centered cubic (fcc) gold lattice was assumed with a lattice constant of $a_0 = 4.078$ \AA. 
A slowly-varying internal field $\bs{u}(\bs{x})$ was generated by first generating uniform random samples for the three components of $\bs{u}(\bs{x})$ ($-0.1 a_0 \leq u_i(\bs{x}) \leq 0.1 a_0$) and then retaining the slow variations by convolving each $u_i(\bs{x})$ with a low-pass filter. 
The crystal spanned $(39, 39, 40)$ voxels along the laboratory frame axes, and therefore a cubic bounding box $\mathcal{V}$ of size $46\times 46\times 46$ voxels was chosen for reconstruction purposes. 
This highly deliberate choice of the bounding box size is for demonstration purposes only. 
In practice it may be chosen based on the size of the Patterson function of the diffraction signal~\cite{Chapman2006}. 
An arbitrary initial lattice orientation was assigned to this digital crystal in the laboratory frame. 
Assuming an x-ray photon energy of $9$ keV, the crystal was actively rotated about $\unitvector{s}_2$ into several Bragg conditions.
The resultant complex-valued object wave was passively transformed into the corresponding detector frame $[\unitvector{k}_1~\unitvector{k}_2~\unitvector{k}_3]$ using the method described in \ref{SS:rotresampl}. 
This two-step composite rotation is denoted by $\mathcal{R}_i$ in Eq.~\eqref{eq.rotate_resampled}. 
The real-space sampling grid of the rotated crystal was then transformed to the span of the basis $\bs{B}_\text{real}^{(i)}$ using the methods described in Sections~\ref{SS:bulk} (bulk) and \ref{SS:shear} (shear).
The rocking direction encoded in the third column of $\bs{B}_\text{recip}^{(i)}$ (\emph{i.e.}, $\bs{q}_k$) was chosen as either a rotation about the $\theta$ or $\phi$ axis (see Figure~\ref{fig.schematic}) based on whether the basis vectors $[\bs{q}_i~\bs{q}_j~\bs{q}_k]$ were maximally orthogonal to each other.
Appendix~\ref{A:MO} describes the customized `mutual orthogonality' (MO) metric used in this choice. For the purpose of this demonstration, five scans were selected based on the highest reciprocal-space MO, corresponding to the $[1\bar{1}\bar{1}]$, $[\bar{1}\bar{1}\bar{1}]$, $[\bar{2}\bar{2}0]$, $[20\bar{2}]$ and $[0\bar{2}\bar{2}]$ lattice planes in their respective Bragg conditions. 
A straightforward 3D FFT of the resampled complex-valued object wave in each case automatically gave the diffracted far-field wave consistent with the correct sampling of a real-world BCDI scan.
The peak intensity for each far-field wave was set to $10^5$ photon counts.
Subsequent addition of Poisson noise to this rescaled diffraction pattern was sufficient to emulate a high-quality signal collected with a synchrotron-based BCDI instrument. 
Thus, the total number of unknowns to solve for was: $4\times 46^3 + 5 = 389349$. 
The prefactor $4$ comes from the electron density $\mathcal{A}$ and the three components of the vector field $\bs{u}(\bs{x})$ at each point $\bs{x}$. 
Further, there are $5$ global scaling factors $\chi_i$ corresponding to each simulated Bragg peak. 
During the MR-BCDI reconstruction, the crystal was initialized to a constant-amplitude cube ($\mathcal{A} = 1$) occupying the entire $46\times 46\times 46$ bounding box with no interior lattice distortion ($\bs{u} = 0$).
The former was achieved by setting $\alpha(\bs{x})$ from Eq.~\eqref{eq.tanh} to 2 and $\alpha_0 = 1$. 
Each of the global scaling factors $\chi_i$ was initialized for the object wave to match the total energy of the corresponding simulated signal.
An Adam optimizer~\cite{Kingma2014} with an initial learning rate of 0.005 was employed. 
A total variation  (TV) regularizer was applied to $\alpha(\bs{x})$ with a coupling constant of $10^{-5}$. 
During optimization, $\bs{u}(\bs{x})$ was constrained to lie within the lattice plane separations $\pm d_{hkl}/2$ corresponding to $[hkl] = [100], [010]$ and $[001]$. 
This constraint may be employed without loss of generality and is discussed further in Section~\ref{S:discuss}.

The optimization was carried out sequentially over increasingly larger `minibatches' of the BCDI data consisting of randomly selected scans, with the final stretch of optimization being performed on all 5 scans. 
Table~\ref{tab:optimplan1} in the Appendix summarizes the sizes of the randomized minibatches and optimization iterations therewith.
After the execution of this plan, the voxels in the crystal interior alone (for which $\mathcal{A} > 0.2$) were optimized for another 1000 iterations over all 5 scans, with the other voxels held constant at their reconstructed values. 
This was done to refine the solution even further within the smaller space consisting of the most relevant optimization dimensions. 
Additionally, after each optimization epoch, a median filter with a kernel size of $3\times 3\times 3$ voxels was applied to the reconstructed lattice distortion field $\bs{u}(\bs{x})$ in order to remove spurious isolated discontinuities induced by phase wraps. 
The median filter was found to be better suited to the task of smoothing $\bs{u}(\bs{x})$ than the TV regularizer used for $\alpha(\bs{x})$. 
The entire optimization took $\sim 2$ hours on an Nvidia Tesla P100 GPU. 
\begin{figure}[ht!]
	\centering
	\includegraphics[width=0.9\textwidth]{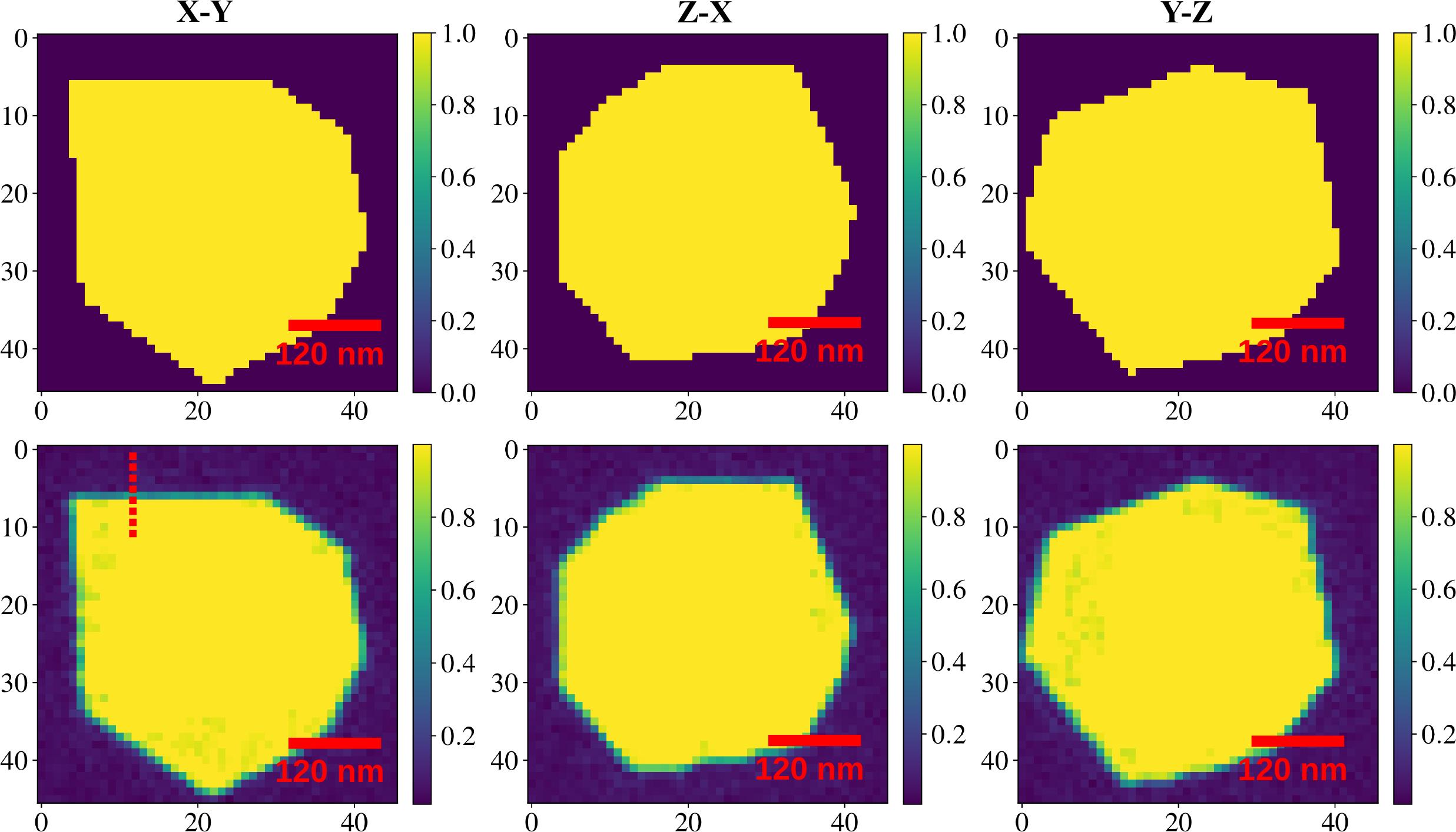}
	\caption{
		Ground truth (\textbf{top}) and reconstructed (\textbf{bottom}) electron density cross-sections ($\mathcal{A}(\bs{x})$) of the synthetic strained crystal with no internal dislocations. 
		The array size represents the original bounding box chosen for reconstruction, \emph{i.e.}, 46 pixels along each lab frame axis. 
		The spatial resolution was estimated by fitting an error function to the reconstructed electron density along the dashed line (see Figure~\ref{fig.nodisloc_metrics}(c)). 
	}
	\label{fig.nodisloc_edens}
\end{figure}
\begin{figure}[ht!]
	\centering
	\includegraphics[width=\textwidth]{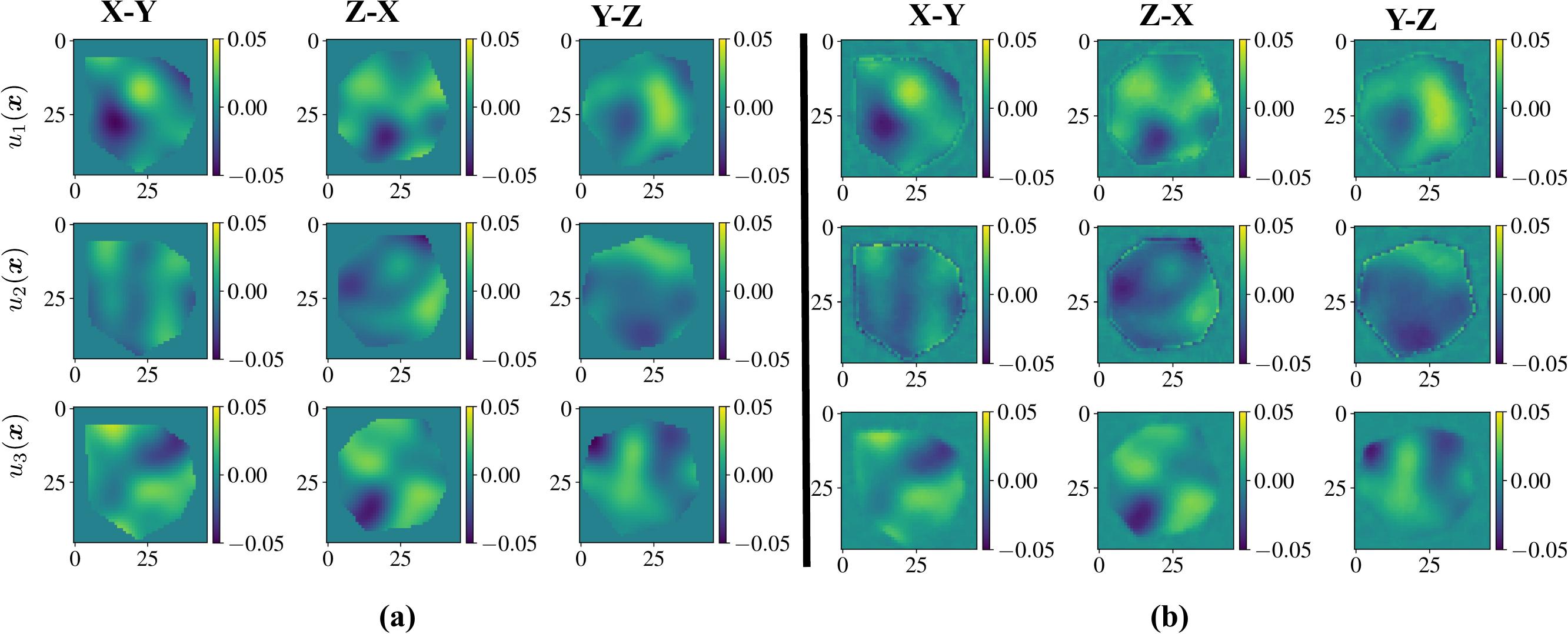}
	\caption{
		Orthogonal cross sections of \textbf{(a)} ground truth and \textbf{(b)} reconstructed lattice distortion components $u_1(\bs{x})$, $u_2(\bs{x})$ and $u_3(\bs{x})$, for the dislocation-free synthetic crystal. 
		The length scale is the same as in Figure~\ref{fig.nodisloc_edens}. 
	}
	\label{fig.nodisloc_u}
\end{figure}
\begin{figure}[ht!]
	\centering
	\includegraphics[width=0.75\textwidth]{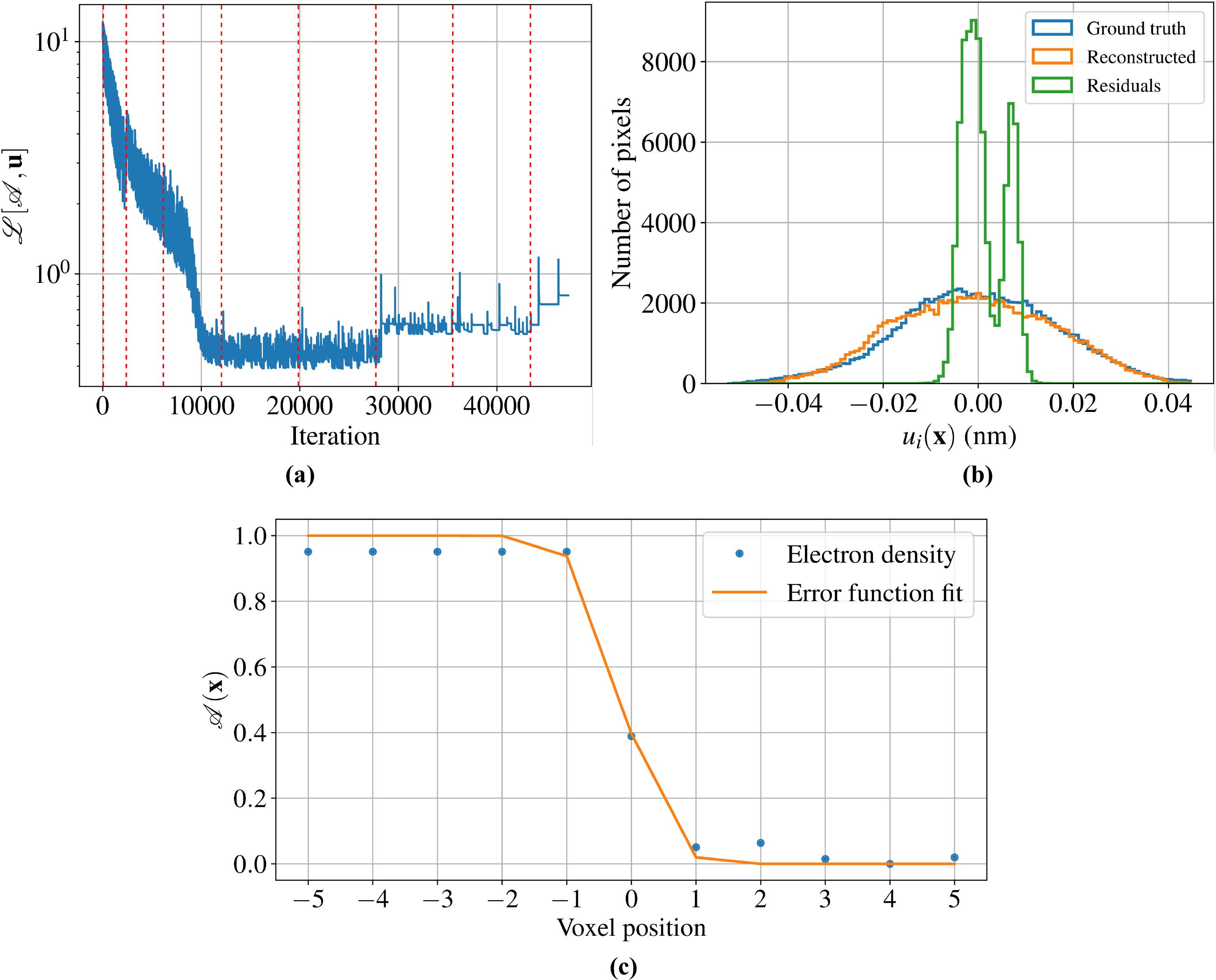}
	\caption{
		\textbf{(a)} Trend in multi-reflection loss function given by Eq.~\eqref{eq.objfun_multi} for the dislocation-free synthetic crystal. 
		The red lines denote the beginning of a new optimization epoch. 
		\textbf{(b)} Histogram of the simulated and reconstructed components of the vector $\bs{u}(\bs{x})$, along with point-to-point residuals of each vector component. 
		\textbf{(c)} Error function fit to the reconstructed profile along the dashed line in Figure~\ref{fig.nodisloc_edens} (X-Y slice). 
		The spatial resolution was estimated from this fit to be $\simeq 1.86$ pixels, or $\simeq 23$ nm. 
	}
	\label{fig.nodisloc_metrics}
\end{figure}
\begin{figure}[ht!]
	\centering
	\includegraphics[width=\textwidth]{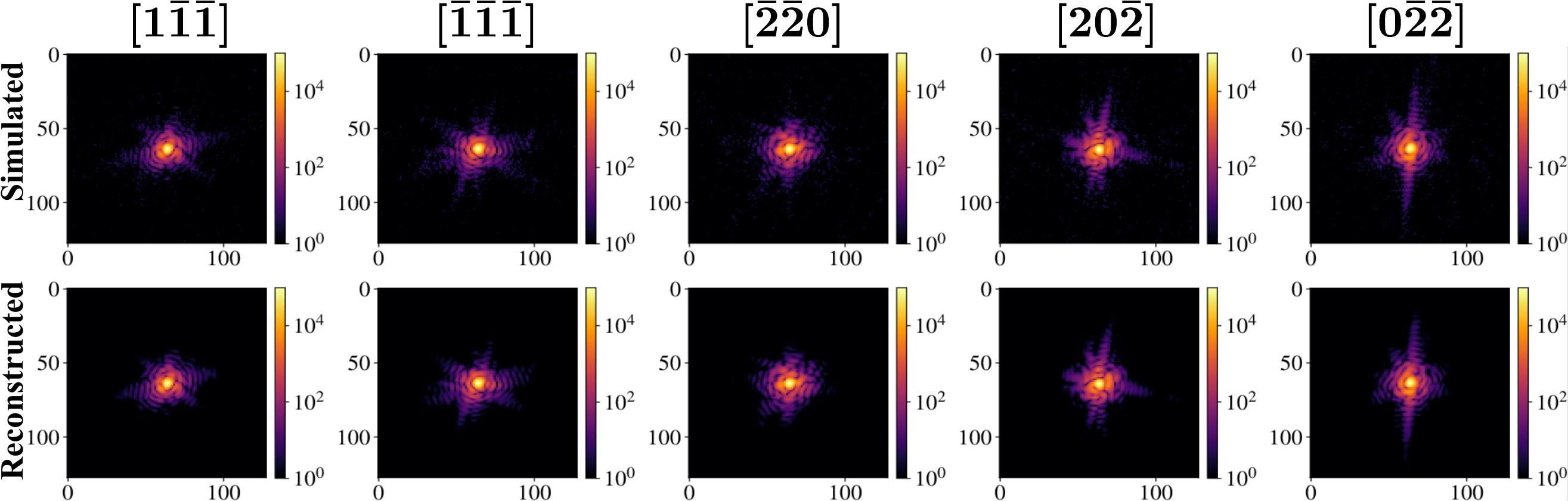}
	\caption{
		Cross-sections of the noisy simulated (top) and reconstructed (bottom) coherent diffraction patterns from the five Bragg peaks chosen for the dislocation-free synthetic crystal. 
		The reconstructed diffraction patterns are shown here without the scaling effect of the $\chi_i$, but rather scaled to match the corresponding simulated intensity peaks. 
		To fully match the color scales, each reconstructed diffraction was clipped below to the smallest nonzero photon count in the corresponding signal (\emph{i.e.}, 1 photon). 
	}
	\label{fig.nodisloc_diffraction}
\end{figure}

Figure~\ref{fig.nodisloc_edens} shows the ground truth and the reconstructed electron density $\mathcal{A}$, on the original grid of step size $s_0 = 12$ nm. 
The spatial resolution of the reconstruction was estimated by fitting an error function to the density profile along the dashed line shown in Figure~\ref{fig.nodisloc_edens}. 
This error function is shown in Figure~\ref{fig.nodisloc_metrics}(c) and has a width of $\sigma = 0.788$ pixels, which is the standard deviation of the underlying Gaussian. 
From this, the spatial resolution was estimated to be $\sqrt{8 \ln 2}\sigma \simeq 1.86$ pixels. 
Figure~\ref{fig.nodisloc_u} shows three orthogonal cross-sections of the three components of $\bs{u}(\bs{x})$ (ground truth and reconstructed). 
Figure~\ref{fig.nodisloc_metrics}(a) shows the progression of the loss function $\mathcal{L}_\text{multi}[\mathcal{A}, \bs{u}]$ over the entire optimization process while Figure~\ref{fig.nodisloc_metrics}(b) shows the histograms of the voxel-by-voxel ground truth and reconstructed vector components of $\bs{u}(\bs{x})$ (all three vector components in the same histogram), as well as the corresponding point-to-point error (residuals). 
The dashed lines in Figure~\ref{fig.nodisloc_metrics}(a) indicate the beginning of each new optimization epoch. 
Within each epoch, the rapid oscillations are attributed to the abrupt change in the loss function landscape due to each new randomized minibatch. 
For example, in the first epoch, the loss function is successively optimized over 400 sets of of 2 randomly selected BCDI scans for 6 iterations each (from Table~\ref{tab:optimplan1}). 
Whenever a new randomized set of scans is used, the energy function landscape abruptly changes from its previous state, and the gradient descent is now, in general, forced along a completely different slope. 

We note that that the majority of the residuals in Figure~\ref{fig.nodisloc_metrics}(b) are concentrated around $0$, and the second peak in residuals originates from the voxels within the margin of mismatch between the estimated object edge with respect to the ground truth.
Prior to calculating these residuals, the reconstruction was spatially aligned with the ground truth by aligning their respective centroids in $\mathcal{A}$, to minimize the effect of any translational offsets. 
The residuals for voxels at which $\mathcal{A} < 0.5$ were ignored in the histogram. 

Figure~\ref{fig.nodisloc_diffraction} shows the comparison of the simulated and reconstructed diffraction patterns. 
The top row depicts the simulated detector measurement at the Bragg peak of each scan. 
The reconstructed diffraction patterns shown here do not contain the scaling effect of their respective $\chi_i$. 
We note the good agreement between the simulated and reconstructed diffraction, which is quantified in Figure~\ref{fig.nodisloc_metrics}(a). 
This shows $\mathcal{L}_\text{multi} \sim 1$, which is the estimate of the aggregate error in signal amplitude per voxel over all the BCDI datasets. 

\begin{figure}[ht!]
	\centering
	\includegraphics[width=\textwidth]{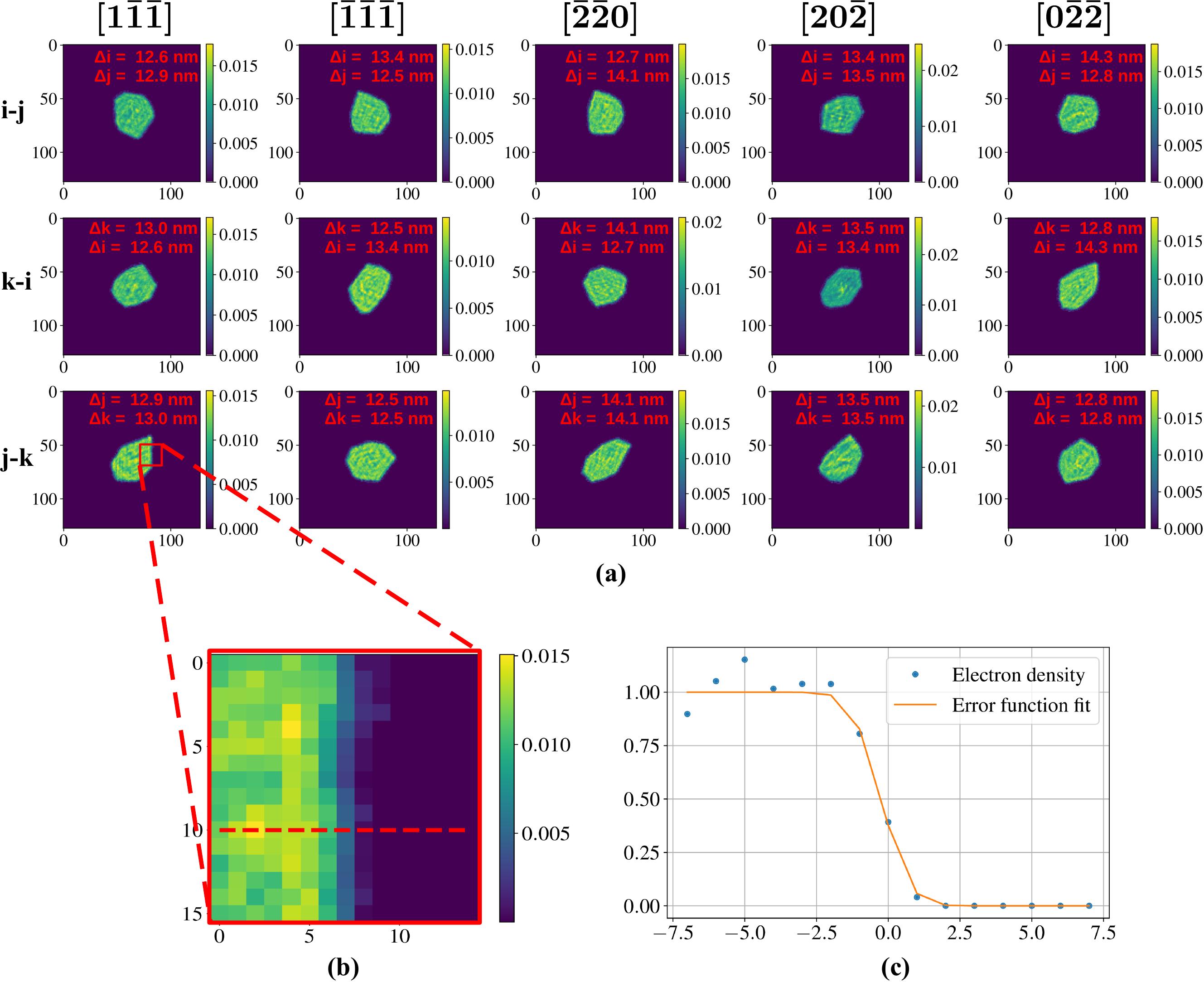}
	\caption{
		\textbf{(a)} 
		Phase retrieval reconstructions of the object wave amplitudes ($\norm{\psi^{(i)}}$), corresponding to the 5 individual BCDI scans from the dislocation-free crystal. 
		The voxel sizes in nanometers are the magnitudes of the columns of the corresponding $\bs{B}_\text{real}^{(i)}$ matrix. 
		We note that these basis vectors are not orthogonal in general, and therefore the slices shown in each column do not correspond to orthogonal slices. 
		\textbf{(b)} 
		Magnified view of one of the reconstructions, with a line profile extracted along the dashed line. 
		\textbf{(c)} 
		Error function fit to the line profile in (b), with an estimated spatial resolution of $1.3\sqrt{8\ln 2} = 3.06$ pixels, or $1.3\sqrt{8\ln 2} \times 13 = 39.8$ nm. 
		The multiplier of $13$ comes for the $\Delta k$ pixel pitch in the corresponding image in (a). 
	}
	\label{fig.nodisloc_pr}
\end{figure}

We now compare these reconstructions to those of conventional phase retrieval. 
Figure~\ref{fig.nodisloc_pr} shows the results of individual phase retrieval reconstructions of the 5 BCDI scans simulated for this crystal (object wave amplitude). 
If the pixels of the final reconstruction $\psi_{ijk}$ in each case are indexed by integers $i$, $j$ and $k$, then each row in Figure~\ref{fig.nodisloc_pr}(a) corresponds to the central slice along the 
$i-j$, $k-i$ and $j-k$ planes. 
These slices are in general not orthogonal, since the real-space sampling directions encoded in each $\bs{B}_\text{real}$ are not orthogonal. 
The phase retrieval was performed with the open-source Python module Phaser~\cite{Maddali2020d}. 
The phase retrieval recipe chosen was a combination of ER, HIO and SF (solvent flipping) with intermittent shrinkwrap-based update of the object support. 
The exact recipe is given in Section~\ref{A:recipe} of the Appendix. 
Each reconstruction took ~0.5 seconds with an Nvidia Tesla P100 GPU. 
An error function fit to the line profile in~\ref{fig.nodisloc_pr}(b) is shown in~\ref{fig.nodisloc_pr}(c), with an estimated spatial resolution of $39.8$ nm. 
We see that the reconstructed electron density from Figure~\ref{fig.nodisloc_edens}, and therefore any subsequent object wave amplitude, is more uniform and exhibits better spatial resolution than those obtained obtained from conventional phase retrieval.

 \subsection{Simulated crystal with screw dislocations} \label{SS:disloc}	A synthetic gold crystal ($a_0 = 4.078$ \AA) with arbitrary facets was generated on the same Cartesian grid with the same voxel size as the dislocation-free case ($s_0 = 12$ nm), with an arbitrary initial crystal orientation. 
The $\bs{u}(\bs{x})$ field at each voxel was consistent with the presence of two spatially separated screw dislocations within the crystal volume, whose cores were along the orthogonal $[111]$ and $[2\bar{2}0]$ crystallographic directions. 
In this paper we favor the well-known continuum model of lattice distortion for screw dislocations~\cite{Phillips2001,Ulvestad2018} for our simulation over the more sophisticated, discontinuity-free atomistic simulations that use known embedded atomic potentials~\cite{Xu2017}. 
The former introduces an unphysical discontinuity in the lattice distortion field, which nevertheless generates the correct outgoing object wave at the desired discretization of $\sim 10$ nm. 
The Burgers vector magnitude $\norm{\bs{b}}$ for each screw dislocation was set to the corresponding lattice-plane spacing. 
Namely, $\norm{\bs{b}}_{111} = a_0/\sqrt{3}$ and $\norm{\bs{b}}_{2\bar{2}0} = a_0/\sqrt{8}$, where $a_0$ is the lattice constant of gold. 
The net displacement field was modeled as the vector sum of the individual displacement fields due to each screw dislocation. 
The simulation procedure of the BCDI data sets, including the selection criterion for Bragg reflections, rocking directions and introduction of Poisson noise, was identical to that for the dislocation-free crystal.
Four peaks with the best reciprocal-space mutual orthogonality (MO) were chosen for reconstruction, corresponding to the $[200]$, $[002]$, $[202]$ and $[2\bar{2}0]$ reciprocal lattice vectors. 
The simulation box $\mathcal{V}$ was chosen to be $40\times 40\times 40$ voxels in size.
The number of optimization variables in this case was therefore $4 \times 40^3 + 4 = 256004$. 
The Adam optimizer was initialized with a learning rate of 0.02, with the chosen sequential optimization plan shown in the Appendix Table~\ref{tab:optimplan2}. 

The final optimization stretch for the interior voxels over all 4 scans was carried out for 5000 iterations. 
As with the dislocation-free crystal, a median filter of size $7\times 7\times 7$ voxels was applied to the components of $\bs{u}(\bs{x})$ after each optimization epoch. 
This optimization took $\sim 3.5$ hours on an Nvidia Tesla P100 GPU. 
\begin{figure}[ht!]
	\centering
	\includegraphics[width=0.9\textwidth]{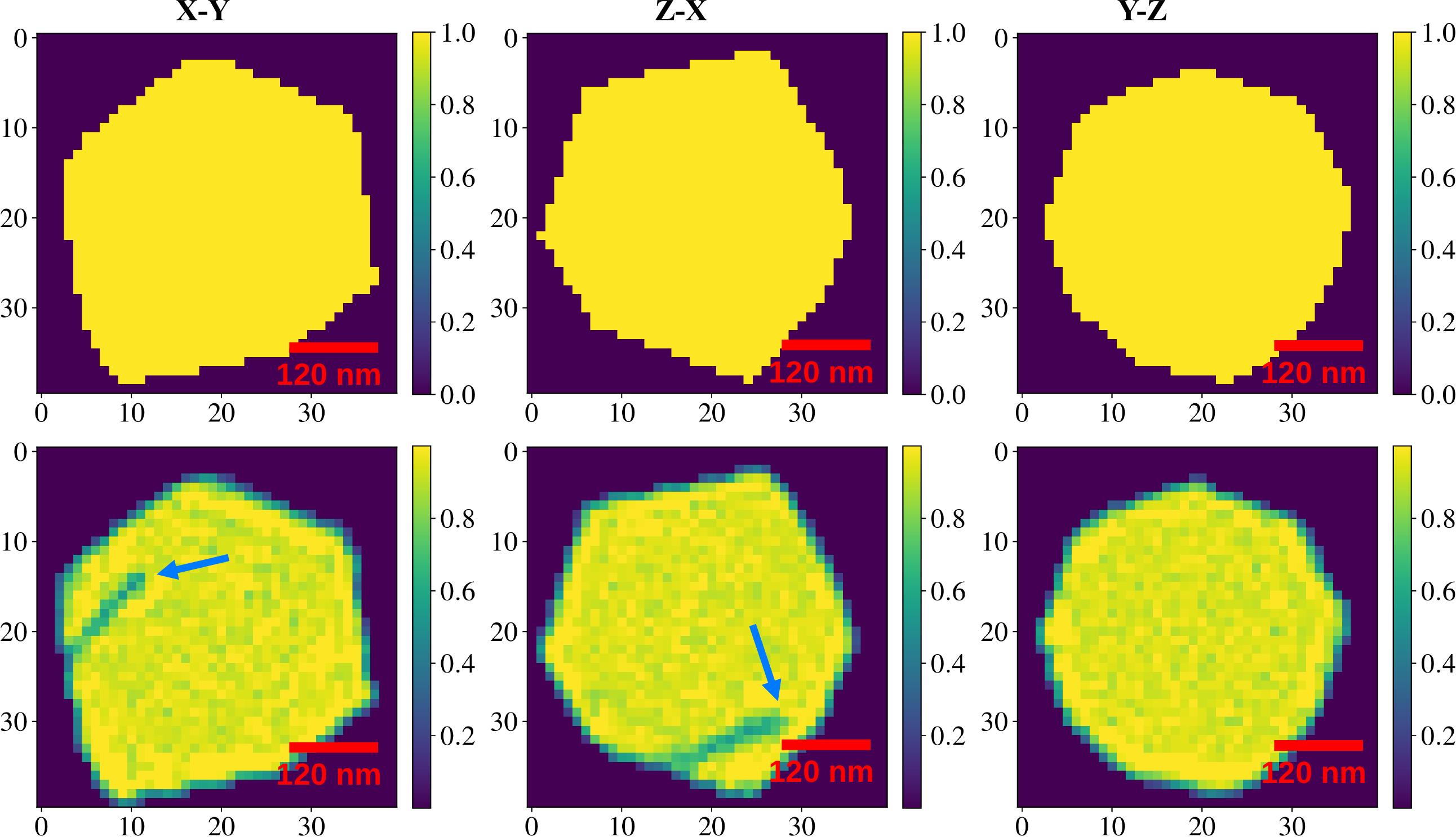}
	\caption{
		Ground truth (\textbf{top}) and reconstructed (\textbf{bottom}) electron density cross-sections of the synthetic strained crystal with two orthogonal screw dislocations. 
		The array size represents the original bounding box chosen for reconstruction, \emph{i.e.}, 40 pixels along each lab frame axis. 
		The arrows indicate the points of intersection of the dislocation cores with the figure plane. 
	}
	\label{fig.screwdisloc_edens}
\end{figure}
\begin{figure}[ht!]
	\centering
	\includegraphics[width=\textwidth]{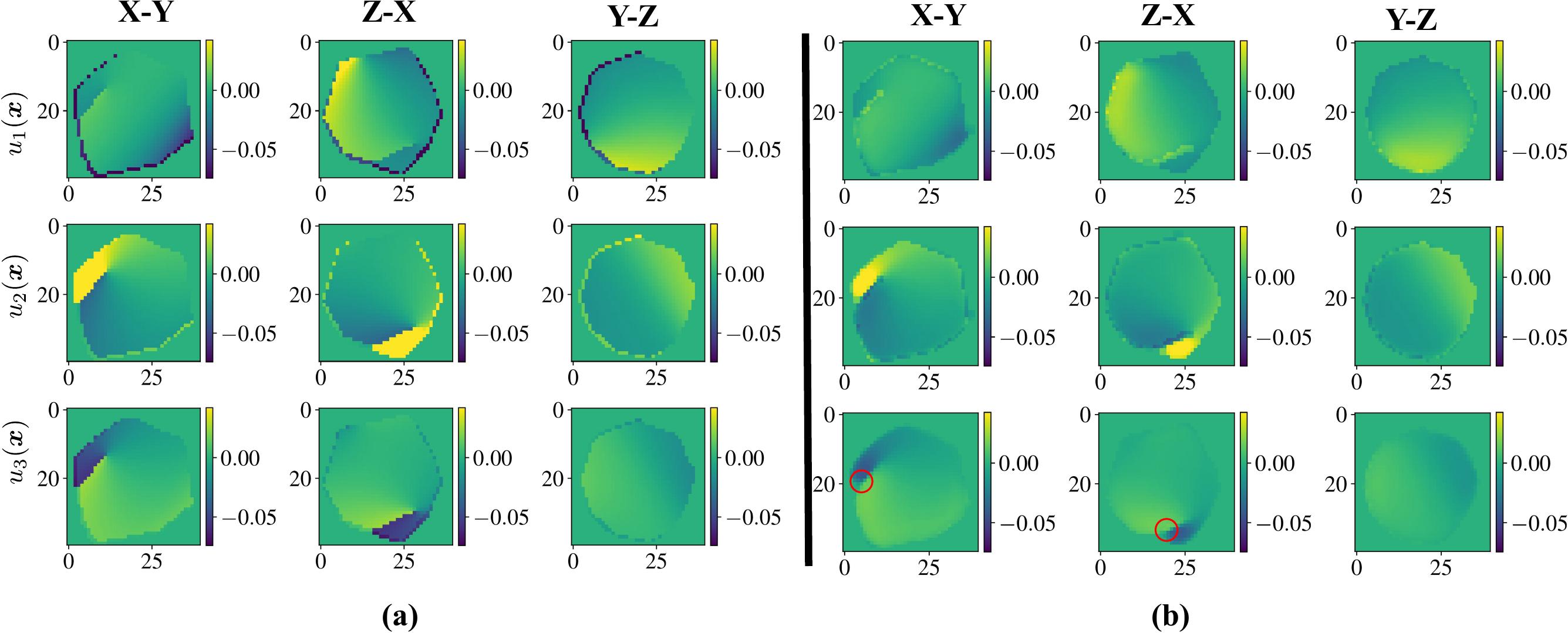}
	\caption{
		Orthogonal cross sections of \textbf{(a)} ground truth and \textbf{(b)} reconstructed lattice distortion components $u_1(\bs{x})$, $u_2(\bs{x})$ and $u_3(\bs{x})$. 
		The length scale is the same as Figure~\ref{fig.nodisloc_edens}. 
		The regions of discontinuity in the lattice distortion match with the drop in electron density in Figure~\ref{fig.screwdisloc_edens}. 
		The circled regions show imperfect reconstruction at locations where the lattice discontinuity meets the crystal surface. 
	}
	\label{fig.screwdisloc_u}
\end{figure}
\begin{figure}[ht!]
	\centering
	\includegraphics[width=\textwidth]{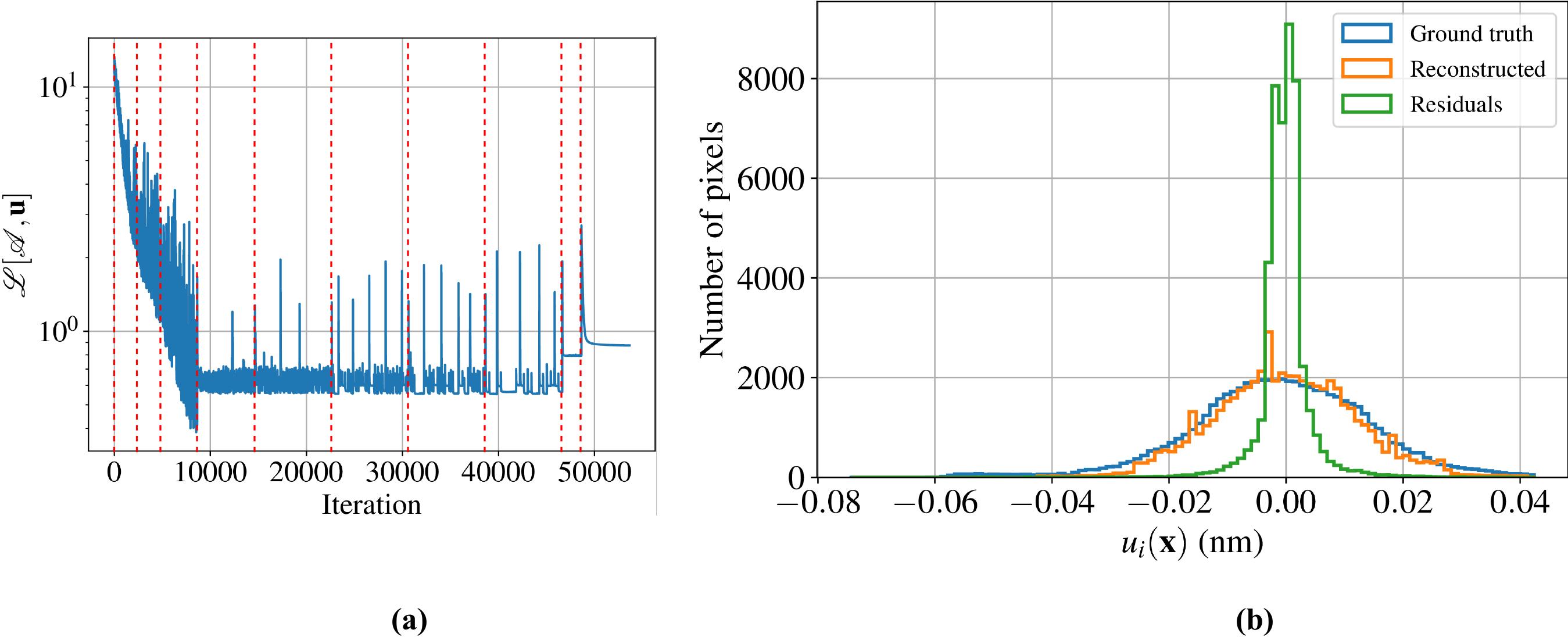}
	\caption{
		\textbf{(a)} Trend in multi-reflection loss function given by Eq.~\eqref{eq.objfun_multi} as a function of Adam optimizer iteration (initial learning rate = 0.02)
		The red lines denote the beginning of a new optimization epoch. 
		\textbf{(b)} Histogram of the simulated and reconstructed components of the vector $\bs{u}(\bs{x})$, along with point-to-point residuals of each vector component. 
	}
	\label{fig.screwdisloc_metrics}
\end{figure}
\begin{figure}[ht!]
	\centering
	\includegraphics[width=\textwidth]{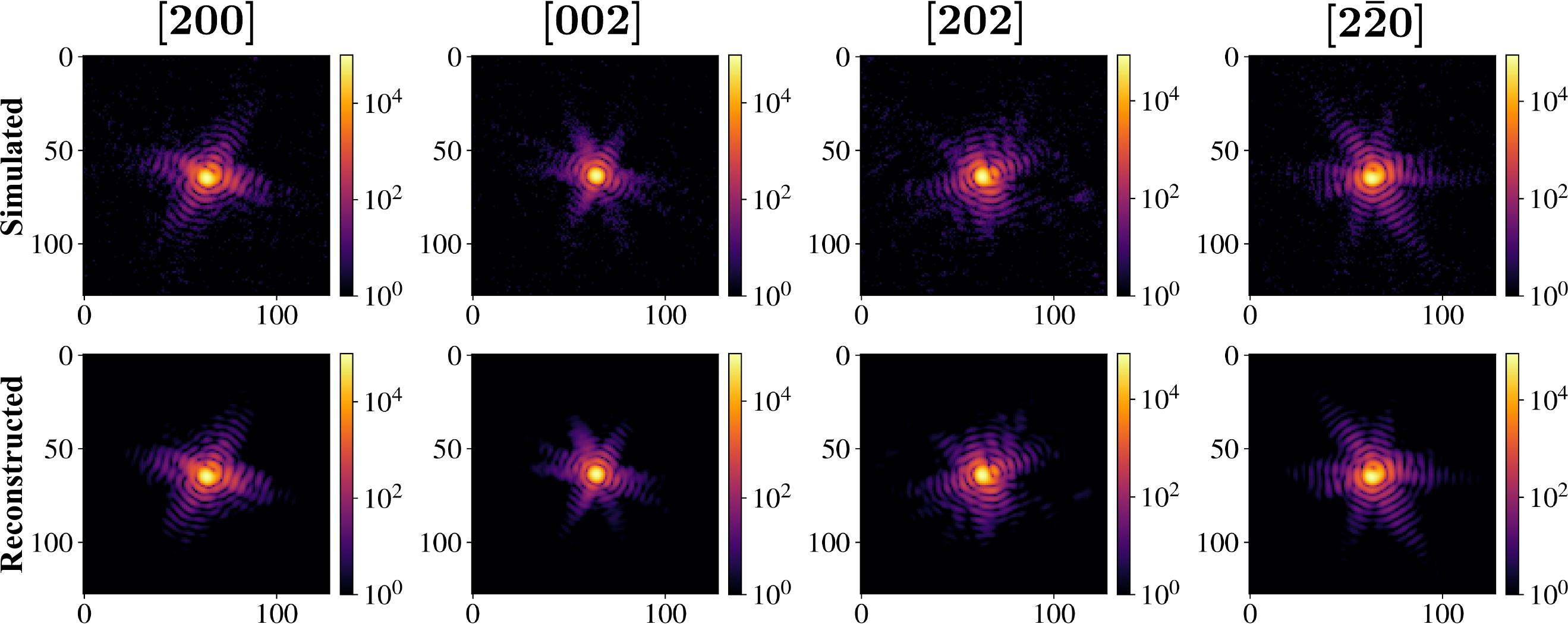}
	\caption{
		Cross-sections of the noisy simulated (top) and reconstructed (bottom) coherent diffraction patterns from the five Bragg peaks chosen for the gold crystal with two orthogonal screw dislocations. 
	}
	\label{fig.screwdisloc_diffraction}
\end{figure}

We again note the good agreement between the simulated and reconstructed crystals (Figures~\ref{fig.screwdisloc_edens} and~\ref{fig.screwdisloc_u}). 
In Figure~\ref{fig.screwdisloc_edens}, we note the drop in reconstructed electron density along the lines where the half-plane of discontinuous lattice distortion intersects the plane of the figure. 
The terminal points of these ``trenches" of low density are the locations of the dislocation cores (marked by arrows).
These trenches are discussed further in Section~\ref{S:discuss}. 
In Figure~\ref{fig.screwdisloc_u} we note the regions circled in red where the discontinuity in the reconstructed $\bs{u}(\bs{x})$ intersects the crystal surface (\emph{i.e.}, an abrupt drop in $\mathcal{A}$) and results in an imperfect reconstruction of $\bs{u}(\bs{x})$. 
Both these effects may be attributed to the inherently band-limited nature of the reconstruction, due to Poisson noise -induced information loss. 

\begin{figure}
	\centering
	\includegraphics[width=\textwidth]{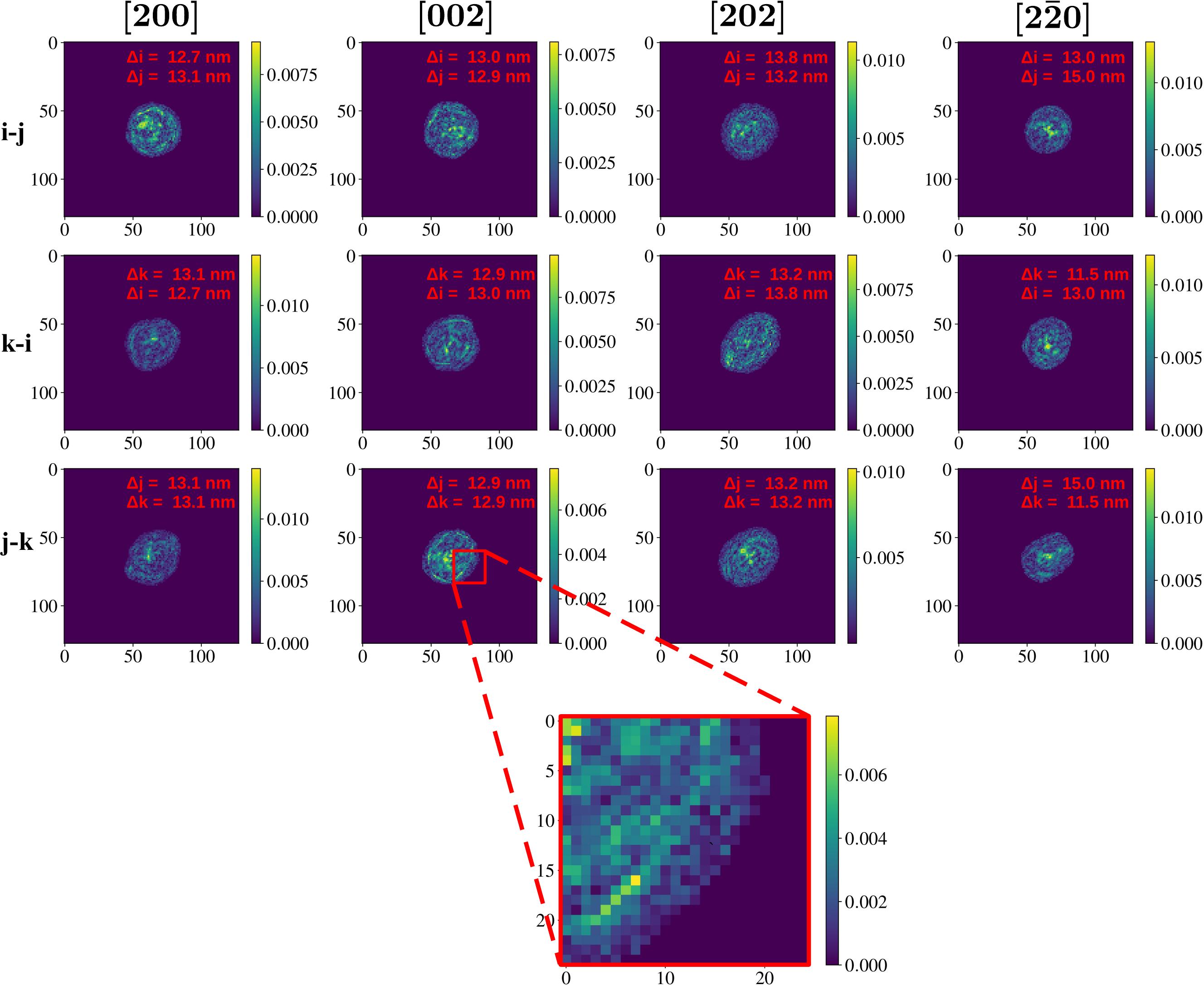}
	\caption{
		\textbf{(a)}
		Individual amplitude reconstructions $\norm{\psi^{(i)}}$ from phase retrieval applied to the BCDI scan the crystal with screw dislocations (complex amplitude), using the recipe in Appendix Section~\ref{A:recipe}. 
		The cross sections shown correspond to the same slices in Figure~\ref{fig.nodisloc_pr}. 
	}
	\label{fig.screwdisloc_pr}
\end{figure}

Figure~\ref{fig.screwdisloc_metrics} shows the reconstruction metrics for this simulated structure similar to those in Figure~\ref{fig.nodisloc_metrics} for the dislocation-free simulated crystal. 
Figure~\ref{fig.screwdisloc_diffraction} compares the simulated and reconstructed diffraction patterns from the four chosen Bragg peaks, on the same color scale. 
Figure~\ref{fig.screwdisloc_pr} shows the individual phase retrieval reconstructions of the BCDI scans from the dislocated crystal lattice. 
The recipe used was the same for the dislocation-free crystal in Figure~\ref{fig.nodisloc_pr}, detailed in the Appendix section~\ref{A:recipe}. 
We note the abject failure of the conventional phase retrieval approach in the presence of dislocations, compared to the reconstruction in Figures~\ref{fig.screwdisloc_edens} and~\ref{fig.screwdisloc_u}. 
Reconstructing dislocated crystals with FPIP algorithms has thus far most successfully been achieved with parallelized reconstruction threads in a guided (\emph{i.e.}, genetic algorithm) approach~\cite{Ulvestad2017a,Hofmann2020}.

 \subsection{Silicon carbide nanocrystal}	 \label{SS:SiC} Having demonstrated MR-BCDI on synthetic datasets, we now move on to experimental data acquired from real crystals. 
We chose silicon carbide (SiC) as a case study, which is a promising platform for quantum sensing and quantum information technologies~\cite{Fuchs2013,Lohrmann2015,Cochrane2016,Atatuere2018,Miao2019,Bourassa2020}. 
The crystal structure is a 4H polymorph with lattice parameters of $a = 3.073$ \AA~and $c = 10.053$ \AA with quasi-hexagonal and quasi-cubic lattice planes. 
Arrays of sub-$\mu$m sized, `D'-shaped SiC crystals were fabricated out of an \emph{i-p-n} SiC wafer procured from Norstel ($[11\bar{2}0]$ major flat, $4^\circ$ miscut from the crystallogrphic surface normal$[0001]$, see Figure~\ref{fig.sic_iso}(d)) and transferred onto a silicon wafer. 
The SiC crystals had known positions, nominally aligned orientations, and separations substantially larger than the x-ray beam footprint. 
This enabled us to collect BCDI data from 6 independent Bragg reflections from an isolated SiC crystal at the 34-ID-C BCDI instrument at the Advanced Photon Source.
These reflections corresponded to the $[10\bar{1}1]$, $[10\bar{1}0]$, $[01\bar{1}2]$, $[01\bar{1}0]$, $[01\bar{1}1]$ and $[10\bar{1}3]$ lattice planes. 
The fabrication process is described in detail in Section~\ref{A:SiC} of the Appendix.
The following pre-processing steps were performed on this experimental data set prior to the MR-BCDI optimization:
\begin{enumerate}
	\item	
		The raw BCDI data from each Bragg reflection was collected as a sequence of $256\times 256$ images with $80$ steps along the rocking curve. 
		The data sets were clipped to a size of $128 \times 128$ pixels in the imaging plane and zero-padded to $128$ steps along the rocking curve to give a cubic simulation volume $128\times 128\times 128$ voxels in size. 
		In addition to reducing memory requirements, this uniform array size ensured that the FFT-based implementation of the shear operator described in Section~\ref{SS:shear} does not suffer from signal normalization issues inherent to applying the FFT along individual array dimensions, while leaving the other dimensions untouched.
		Future iterations of this reconstruction tool will be generalized to non-uninform array dimensions. 
\item
Unlike with the simulations, the precise lab frame crystal orientation was known only approximately prior to reconstruction and had to be refined. 
		This was done by a simple least squares optimization of the crystal orientation (expressed as  a 3-parameter rotation vector) over the $\delta$, $\gamma$ and $\theta$ motor positions (see Figure~\ref{fig.schematic}) of all 6 BCDI scans with the Nelder-Mead optimizer~\cite{Nelder1965} available in the SciPy software package. 
\end{enumerate}
The maximum pixel intensities in each of the 6 measured Bragg peaks were respectively observed to be: 40980, 16072, 32661, 14577, 34123 and 7805 photons. 

\begin{figure}[ht!]
	\centering
	\includegraphics[width=0.4\textwidth]{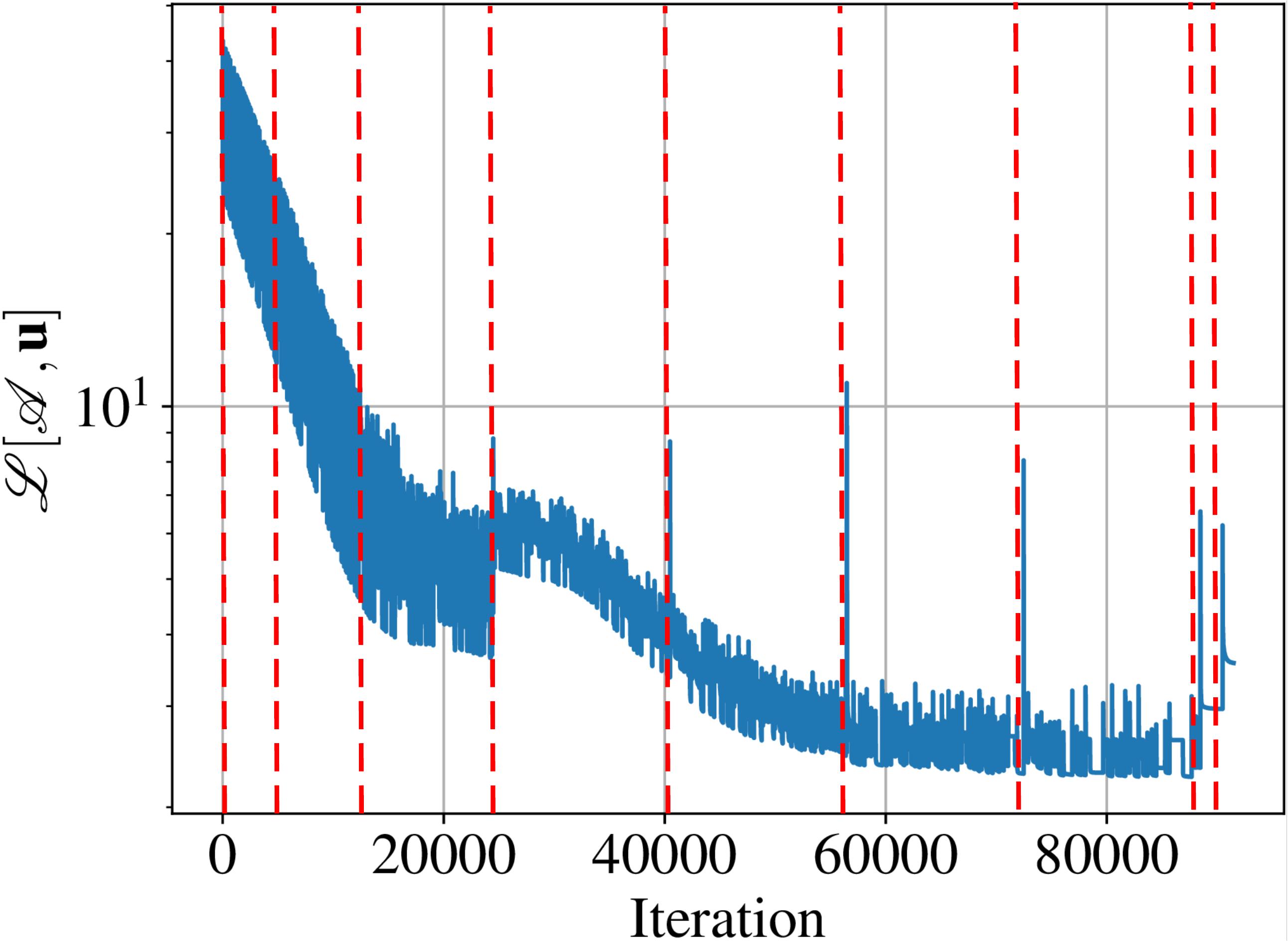}
	\caption{
		Loss function trend for the SiC crystal reconstruction, for the optimization plan shown in Table~\ref{tab:optimplan3}. 
	}
	\label{fig.sic_error}
\end{figure}
\begin{figure}[ht!]
	\centering
	\includegraphics[width=0.9\textwidth]{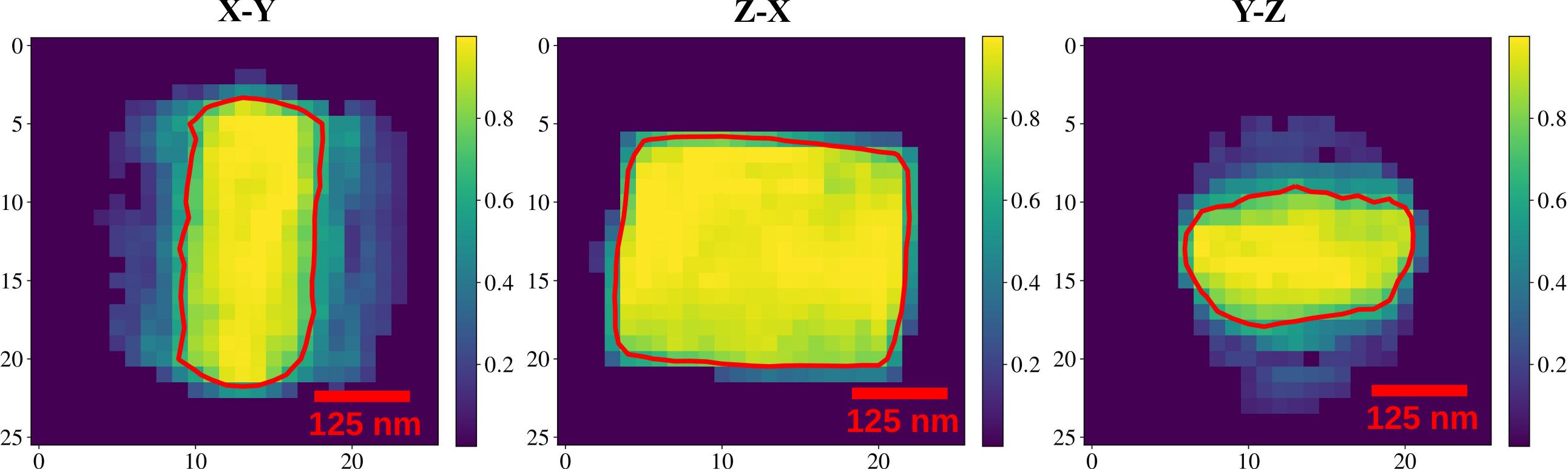}
	\caption{
		Final estimated SiC nanocrystal electron density $\mathcal{A}$, from a highly permissive mask threshold of $0.1$. 
		The red line is the contour at a threshold of $\mathcal{A} = 0.65$, representing the approximate crystal surface. 
	}
	\label{fig.sic_edens}
\end{figure}
\begin{figure}[ht!]
	\centering
	\includegraphics[width=0.7\textwidth]{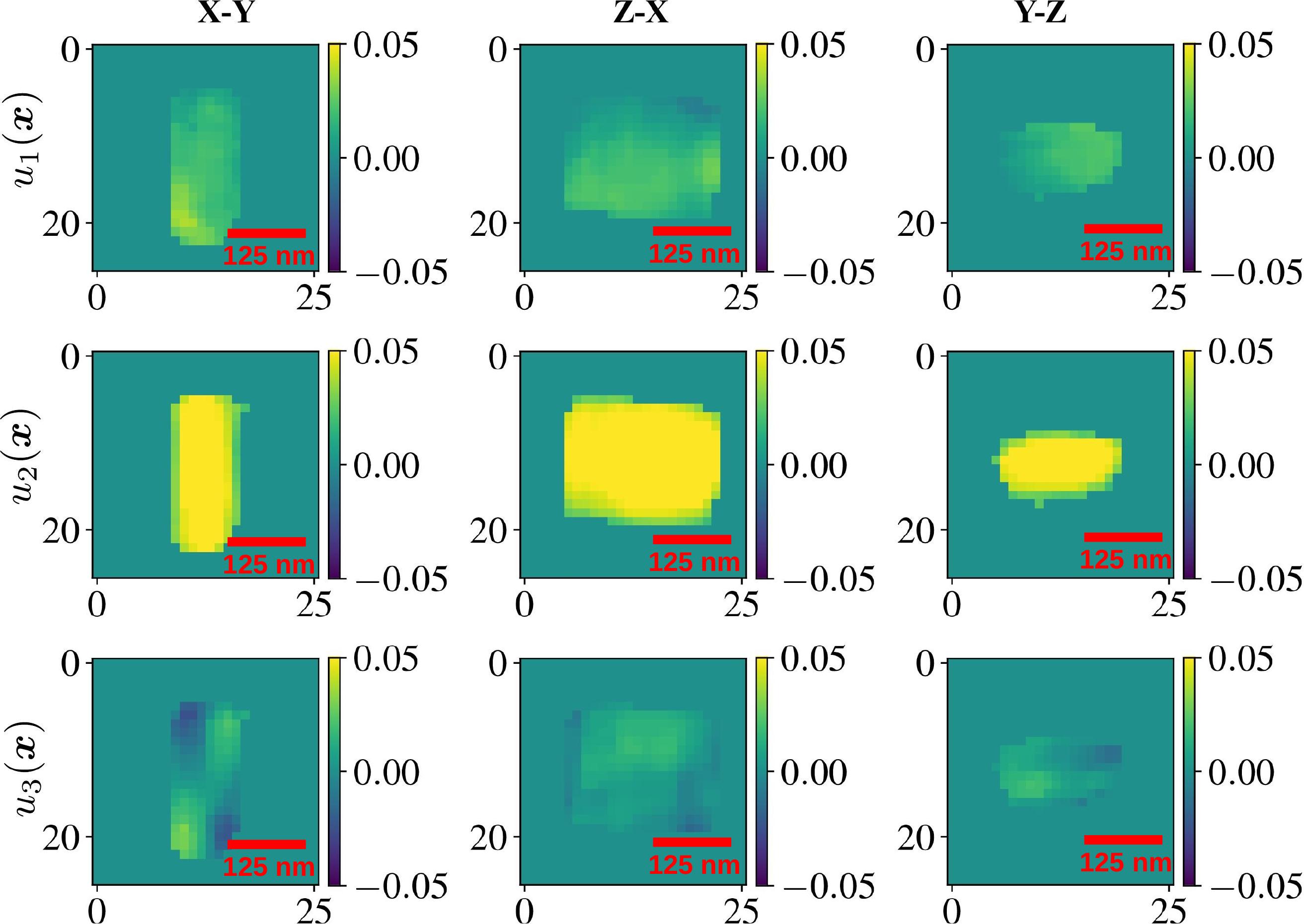}
	\caption{
		Reconstructed components of the lattice distortion within the SiC crystal (centered at zero volume-averaged distortion in each dimension), within the estimated crystal surface at $\mathcal{A} = 0.65$. 
}
	\label{fig.sic_u_recon}
\end{figure}
\begin{figure}[ht!]
	\centering
	\includegraphics[width=0.8\textwidth]{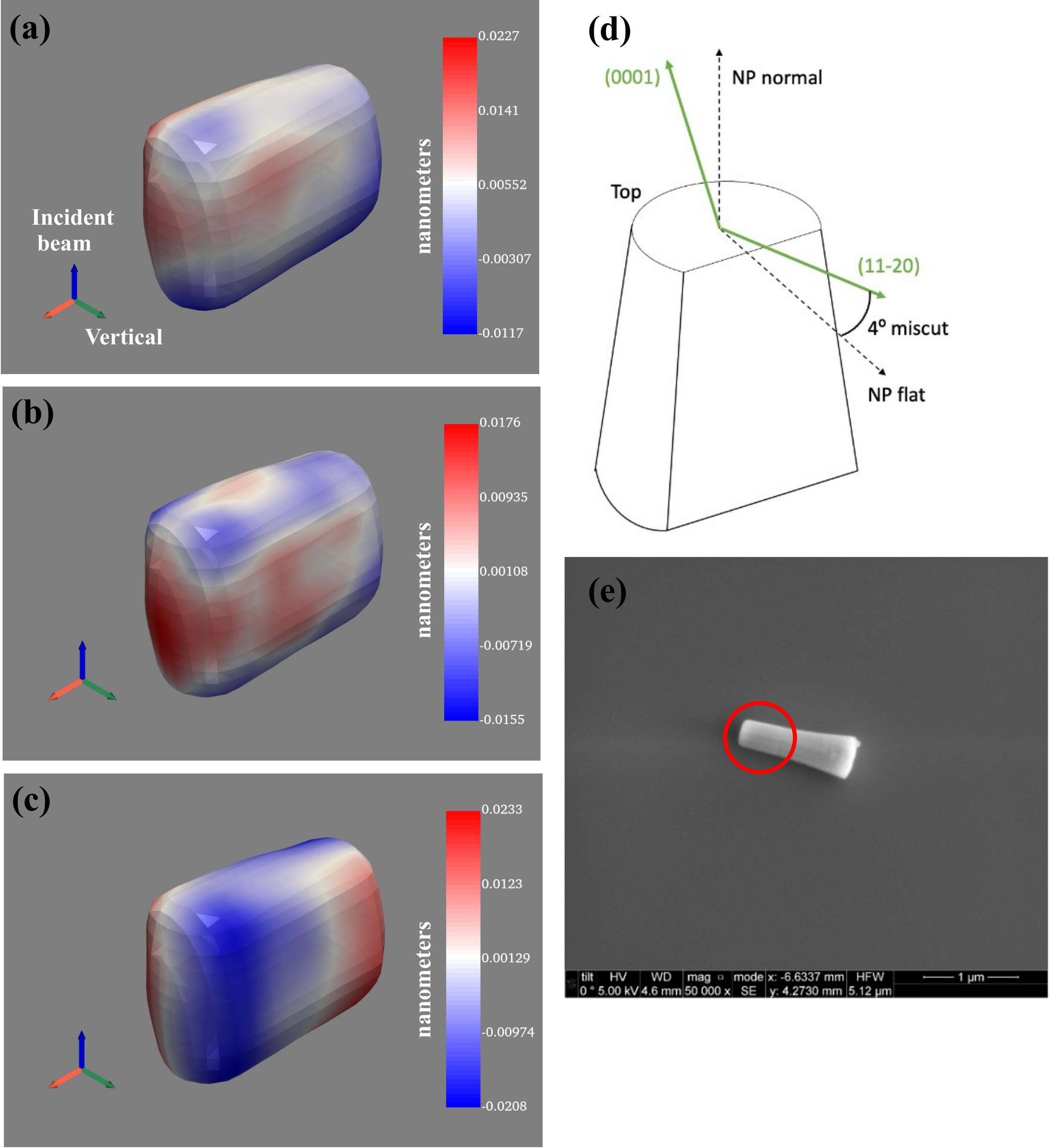}
	\caption{
		\textbf{(a), (b), (c)} Isosurface plots of the reconstructed $u_1(\bs{x})$, $u_2(\bs{x})$ and $u_3(\bs{x})$ respectively in the SiC crystal (color scale in nanometers), with an isovalue of $0.65$ in electron density.			
		\textbf{(d)} Schematic of the nanoparticle, with the `D'-shaped frustum cross section, also visible in the cross section in Figure~\ref{fig.sic_edens}(c). 
		Also indicated are the $[0001]$ and $[11\bar{2}0]$ crystallographic planes relative to the facets. 
		\textbf{(e)} SEM image of the nanoparticle structure on the substrate, with the diffracting SiC portion highlighted within the red circle.  
	}
	\label{fig.sic_iso}
\end{figure}
\begin{figure}[ht!]
	\centering
	\includegraphics[width=\textwidth]{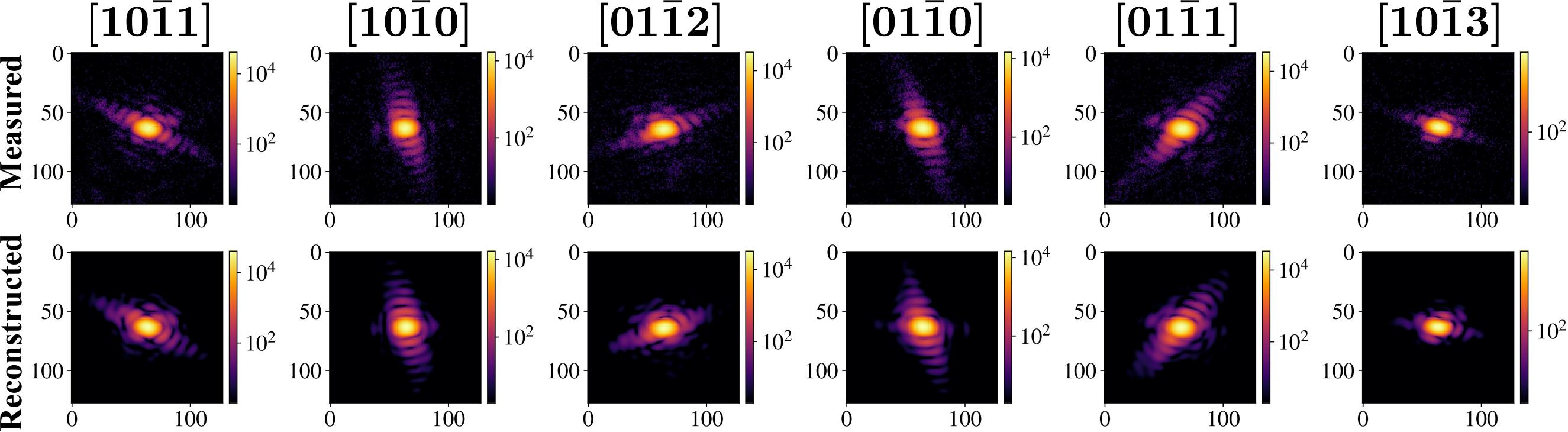}
	\caption{
		Measured (\textbf{top}) and reconstructed (\textbf{bottom}) diffraction patterns of SiC. 
	}
	\label{fig.sic_diffraction}
\end{figure}

Figure~\ref{fig.sic_error} depicts the progression of the loss function for the optimization scheme described in Table~\ref{tab:optimplan3}. 
Figure~\ref{fig.sic_edens} shows the electron density after optimization of the interior voxels, `interior' being defined in this case as above the highly permissive threshold of $\mathcal{A} > 0.1$. 
Also shown is the contour at $\mathcal{A} = 0.65$, depicting the approximate surface of the crystal.  
Figure~\ref{fig.sic_u_recon} shows the reconstructed components of $\bs{u}(\bs{x})$ along with the same contour. 
Figure~\ref{fig.sic_iso} shows the corresponding 3D isosurface plots of the components of $\bs{u}$, with the color scale denoting the spatially varying lattice distortion in nanometers. 
We see from the absence of discontinuities that the particle appears to lack dislocations or sharply varying strain fields.
Figure~\ref{fig.sic_diffraction} shows the measured and inferred diffraction patterns, with the reconstructed patterns transformed to the ranges of the measured signals. 

We see that the known `D'-shaped cross-section of the nanoparticle in Figure~\ref{fig.sic_iso}(d) is discernable in the electron density and lattice distortion reconstructions shown in Figures~\ref{fig.sic_edens} and~\ref{fig.sic_u_recon}, despite the loss of spatial resolution as a result of clipping the BCDI data sets to a $128\times 128\times 128$ -pixel array.  
The relatively uniform frustum of the nanocrystal in the~\ref{fig.sic_iso}(a), (b) and (c) isosurface plots is corroborated by the SEM image in~\ref{fig.sic_iso}(e). 
We note that the SEM image shows the highly porous \emph{p}-type tail from the fabrication process. 
This volume did not diffract appreciably compared to the \emph{i}-type section, so only the latter was reconstructed. 
  \section{Discussion}											\label{S:discuss}	In this paper we present a completely differentiable forward model for the MR-BCDI reconstruction problem for vector-valued lattice distortion fields, tailored to non-uniformly strained and defective crystalline nanoparticles. 
We have shown reconstruction results from simulated and experimental coherent diffraction data. 
By using a gradient descent-based global optimization over randomized minibatches, we have obviated the need for trial-and-error FPIP recipes and, more significantly, support update via the shrinkwrap algorithm.  
We have demonstrated consistently better reconstructions for both defect-free and defected crystals with dislocations that are isolated on the scale of BCDI resolution. 
The reconciliation of different signal gridding geometries with the Fourier transform is crucial for reliable imaging of such topologically intricate structural heterogeneities within the crystalline nanoparticles.
This is particularly true for heterogeneities that manifest nominally as phase wraps and discontinuities of the coherent diffracted wave, such as edge and screw dislocations. 
We have demonstrated that BCDI data sets from independent Bragg reflections, when of sufficiently high quality and explicitly coupled together, can `conspire' to give a more uniform electron density profile and potentially better spatial resolution (as evidenced by the estimated $23$ nm and $39.8$ nm spatial resolutions for the respective MR-BCDI and phase retrieval reconstructions of the dislocation-free crystal). 
This is consistent with previous experimental findings that data redundancy from coupled diffraction data sets improves spatial resolution in  BCDI~\cite{Wilkin2021}. 

On the topic of shrinkwrap~\cite{Marchesini2003}, we note that this currently prevalent method is essentially a `blind' process contingent on reasonable prior convergence to a compact shape during phase retrieval. 
Failing this, it becomes irrelevant and potentially misleading, as our examples with the screw dislocations indicate (Figure~\ref{fig.screwdisloc_pr}). 
Our global optimization approach avoids this inconvenience entirely by simply making the support one of the products of the fitting process. 
The only choices involved are the size of the initial bounding box $\mathcal{V}$, and the hyperparameter $\alpha_0$ from Eq.~\eqref{eq.tanh}. 
In contrast, shrinkwrap involves two hyperparameters, namely the width of the blurring Gaussian kernel (usually denoted $\sigma$) and an arbitrary threshold below which the amplitude is suppressed to zero. 
In addition, there is the more subtle choice of how often to employ shrinkwrap in a recipe, a largely trial-and-error process when reconstructing heavily dislocated crystals. 

This formulation of MR-BCDI offers another potential advantage in that it permits the analytical interrogation of the high-dimensional solution space in the neighborhood of the true solution ($4\times 46^3 + 5 = 389349$ dimensions in the case of our dislocation-free synthetic crystal). 
Without the explicit evaluation of a loss function, FPIP methods offer little to no insight in this regard, preventing rigorous BCDI error analysis. 
In contrast, our analytical prescription is a potential first step in developing rigorous error propagation and uncertainty quantification capabilities, which is currently lacking in current FPIP-based BCDI and will be in high demand as expected improvement in data quality from fourth-generation synchrotrons pushes the limits of spatial resolution. 

In Section~\ref{SS:nodisloc} we noted that we restricted $\bs{u}(\bs{x})$ to lie within the lattice plane spacings defined by the $[100]$, $[010]$ and $[001]$ directions of the crystal. 
These directions are not orthogonal in general, as is seen from the equivalent $[10\bar{1}0]$, $[01\bar{1}0]$ and $[0001]$ for our SiC crystal, in which the last two directions are separated by $120^\circ$. 
This constraint is not a loss of generality when we consider the conditions under which phase wrapping takes place for a particular Bragg reflection. 
The phase at a point $\bs{x}$ corresponding to a reflection $\bs{G}_{hkl}$ is given by: $\phi(\bs{x}) \equiv 2\pi \bs{G}_{hkl} \cdot \bs{u}(\bs{x})$. 
Therefore a phase wrap is associated with an excess displacement $\bs{d}_{hkl}$ given by: $\phi(\bs{x}) + 2m\pi = 2\pi \bs{G}_{hkl} \cdot [ \bs{u}(\bs{x}) + m\bs{d}_{hkl} ]$, where $m \in \mathbb{Z}$. 
		Here, $\bs{d}_{hkl}$ is the vector in the direction of $\bs{G}_{hkl}$ and whose magnitude is equal to the $hkl$ lattice plane spacing, with $\norm{\bs{G}_{hkl} \cdot \bs{d}_{hkl}} = 1$ in crystallographers' units ($2\pi$ in physicists' units). 
We see that for a given $\bs{G}_{hkl}$, $\bs{u}(\bs{x})$ is determined up to an integer multiple of $\bs{d}_{hkl}$ in the same sense that the phase is only determined up to an integer multiple of $2\pi$. 
		We therefore deduce that any $\bs{u}(\bs{x})$ may be reduced to a `fundamental zone' (terminology borrowed from crystallography) between the parallel lattice planes located at $\bs{u}(\bs{x}) = \bs{0}$ and $\bs{u}(\bs{x}) = \bs{d}_{hkl}$ (or equivalently, $\pm \bs{d}_{hkl}/2$), by addition or subtraction of an appropriate number of $\bs{d}_{hkl}$~\cite{Kandel2021a}. 
The reduced $\bs{u}(\bs{x})$ is guaranteed to generate the same phase in the outgoing object wave. 
		Thus, as a purely computational convenience, we constrain $\bs{u}(\bs{x})$ at each point to lie within the intersection of the three largest `fundamental zones' of the crystal lattice, which correspond to the three largest (close packed) lattice plane separations (\emph{i.e.}, $\norm{\bs{d}_{100}}$, $\norm{\bs{d}_{010}}$ and $\norm{\bs{d}_{001}}$). 
This volume of intersection is rectangular for a cubic or orthorhombic crystal, but in general is a sheared parallelopiped. 
This constraint serves the following purposes: 
\begin{enumerate}
	\item
		It plays a role in preventing spurious discontinuities in the reconstructed $\bs{u}(\bs{x})$ resulting from excessively large gradient descent steps, which could cause the $\bs{u}(\bs{x})$ at adjacent voxels to converge to different but phase-equivalent values.
	\item
		It retains faithful representation of the \emph{genuine} discontinuities in $\bs{u}(\bs{x})$ from lattice defects that are equal in size to a closed-packed lattice plane spacing, such as edge and screw dislocations. 
\end{enumerate}

We also note that the conjugate basis formalism of $\br$ and $\bq$ is \emph{not} the most general parameterization of the signal space. 
In fact, it immediately breaks down when imaging sufficiently small crystals, for which sufficient fringe information is obtained over a much larger volume of Fourier space, whose size is comparable to $\norm{\bs{G}_{hkl}}$. 
Under these circumstances, the curvature of the Ewald sphere can no longer be ignored when rocking the sample. 
Accounting for this aspect would require the design of better forward models, which is currently an active field of study. 
Further, the use of the fast Fourier transform (FFT)  leaves us open to the problem of cyclic aliasing of the inferred scattering. 
This imposes a limit on the spatial resolution of MR-BCDI that is not consistent with the amount of Fourier-space information collected in a scan. 
Perversely, this effect is more pronounced with a higher signal-to-noise ratio measurement, which is expected at a fourth-generation synchrotron.
Research to address this issue is ongoing in the BCDI community~\cite{Godard2021}. 

All this taken into account, this work represents a significant step in the imaging of defected materials on the  $\sim 10$ nm scale, while eliminating the need for \emph{ad hoc} phase retrieval recipe development and shrinkwrap-based support updates. 
In addition, it is another step in bringing coherent diffraction methods into the scope of global optimization techniques~\cite{Kandel2019,Kandel2021}, thereby enabling the use of highly optimized software packages capable of handling large data volumes and running on seemingly ever-improving high-performance computing hardware. 

 \section{Methods: Fourier transform-based resampling}		\label{S:ftresmpl}	We start with the discrete object $\psi(s_0\bs{m})$, which denotes the object $\psi(\bs{x})$ sampled along the orthogonal directions $[\unitvector{k}_1~\unitvector{k}_2~\unitvector{k}_3]$ of the detector frame, in steps of size $s_0$.
We assume the digitized scatterer to be inside a cubic array of size $N_1\times N_2\times N_3$ pixels. In the MR-BCDI problem, this is one of $i$ mounting configurations corresponding to a single Bragg condition. 
We now describe the basic operations of (i) bulk deformation, (ii) shear deformation and (iii) rigid-body rotation of the sampling grid.

The basis matrix $\bs{B}_\text{real}$ for this demonstrative configuration is based on an actual BCDI measurement described in Ref.~\cite{Maddali2020a}:
\begin{equation}\label{eq.breal}
	\bs{B}_\text{real} = \left[
		\begin{matrix}
			12.72 & 0 & 0 \\
			0 & 12.72 & 0 \\
			3.21 & 2.70 & 11.87
		\end{matrix}
	\right]~\text{nanometers}
\end{equation}

The resampling method $\psi(s_0\bs{m}) \rightarrow \psi(\bs{B}_\text{real}\bs{m})$ consists of two interpolation operations acting on $\psi(s_0\bs{m})$, performed in order: 
(i) `bulk' resampling along the orthogonal directions in \emph{new} steps of $B_{11}$, $B_{22}$ and $B_{33}$, where $B_{ii}$ denote the diagonal elements of $\bs{B}_\text{real}$, followed by (ii) a `shear' resampling into each of the three independent orthogonal directions, corresponding to the off-diagonal elements of $\bs{B}_\text{real}$.
We introduce for simplicity $\mathcal{F}_i$ as the 1-D Fourier transform along the axis $i$ (where $i = 1, 2, 3$). 

Further, we assign $s_0 = (\det \bs{B}_\text{real})^{1/3}= 12.43$ nm. 
This choice allows us to demonstrate both bulk stretching and compression, since is it greater than one of the $B_{ii}$ but less than the other two. 
For an MR-BCDI experiment, $s_0$ is the size of the reconstruction grid within the bounding box $\mathcal{V}$. 

\subsection{Bulk resampling}				\label{SS:bulk}			Real-space resampling along the orthogonal array axes of $\psi(s_0\bs{m})$ is achieved by truncating or padding the numerical array in Fourier space. 
To this end, we define operations $\mathcal{P}$ and $\mathcal{T}$ that symmetrically pad or truncate the input array along the specified axis: 
\begin{itemize}
	\item	
		$\mathcal{P}(\unitvector{k}_i, n)\psi_{\bs{m}}$ denotes the operator $\mathcal{P}$ acting on an array $\psi_{\bs{m}}$ that returns an array zero-padded on either side by $n$ pixels. 
		This results in $2n$ more pixels along the $\unitvector{k}_i$ axis.
	\item	
		$\mathcal{T}(\unitvector{k}_i, n)\psi_{\bs{m}}$ denotes the operator $\mathcal{T}$ acting on an array $\psi_{\bs{m}}$ that returns an array truncated/clipped on either side by $n$ pixels. 
		This results in $2n$ fewer pixels along the $\unitvector{k}_i$ axis.
\end{itemize}

We seek to resample the real-space object array $\psi_{\bs{m}}$ along axis $\unitvector{k}_i$ in steps of $B_{ii}$ (different from the original sampling steps size $s_0$). 
The number $n$ of pixels by which to zero-pad/truncate is easily computed from the relation between the real-space step size and the Fourier-space numerical aperture: $s_0 \propto 1/N$. 
If $N_i^\text{(new)}$ is the number of pixels along $\unitvector{k}_i$ after zero-padding/truncation, then we have: 
\begin{align}
	N_i^\text{(new)} &= \left(\frac{s_0}{B_{ii}} \right) N_i \\
	\Longrightarrow n &= \frac{1}{2} \left\lvert N_i^\text{(new)} - N \right\rvert
		= \frac{1}{2}\left\lvert N\left(1 - \frac{s_0}{B_{ii}}\right)\right\rvert
		\label{eq.ncalc}
\end{align}
Here $\left\lvert \cdot \right\rvert$ denotes the absolute value and the factor $1/2$ comes from the requirement to pad/truncate symmetrically. 
Armed with this notation, the resampling operator $\bulkresample{i}{}$ is given by the following sequence of operations:
\begin{equation}
	\bulkresample{i}{} = \left\{
		\begin{matrix}
			\resamplefine{\unitvector{k}_i}{n}{i} & \text{if} & B_{ii} < s_0 \\
			\resamplecoarse{\unitvector{k}_i}{n}{i} & \text{if} & B_{ii} > s_0
		\end{matrix}
	\right.
	\label{eq.bulkresampl}
\end{equation}
The bulk-resampled array is denoted as: $\psi^\prime_{\bs{m}} = \bulkresample{i}{}\psi_{\bs{m}}$. 
The value of $n$ in Eq.~\eqref{eq.bulkresampl} is computed from Eq.~\eqref{eq.ncalc}, in practice rounded to the nearest integer. 
The final, bulk-resampled object along all three orthogonal axes is given by: 
\begin{align}
	\psi^\prime_{\bs{m}} &\equiv \psi([B_{11}~B_{22}~B_{33}]\bs{m}) \notag \\ 
		&= \underbrace{
			\bulkresample{1}{}\bulkresample{2}{}\bulkresample{3}{}
		}_\text{total bulk resampling operator}
		\psi_{\bs{m}}
	\label{eq.bulkresmpl_full}
\end{align}

\begin{figure}\centering
	\includegraphics[width=0.65\textwidth]{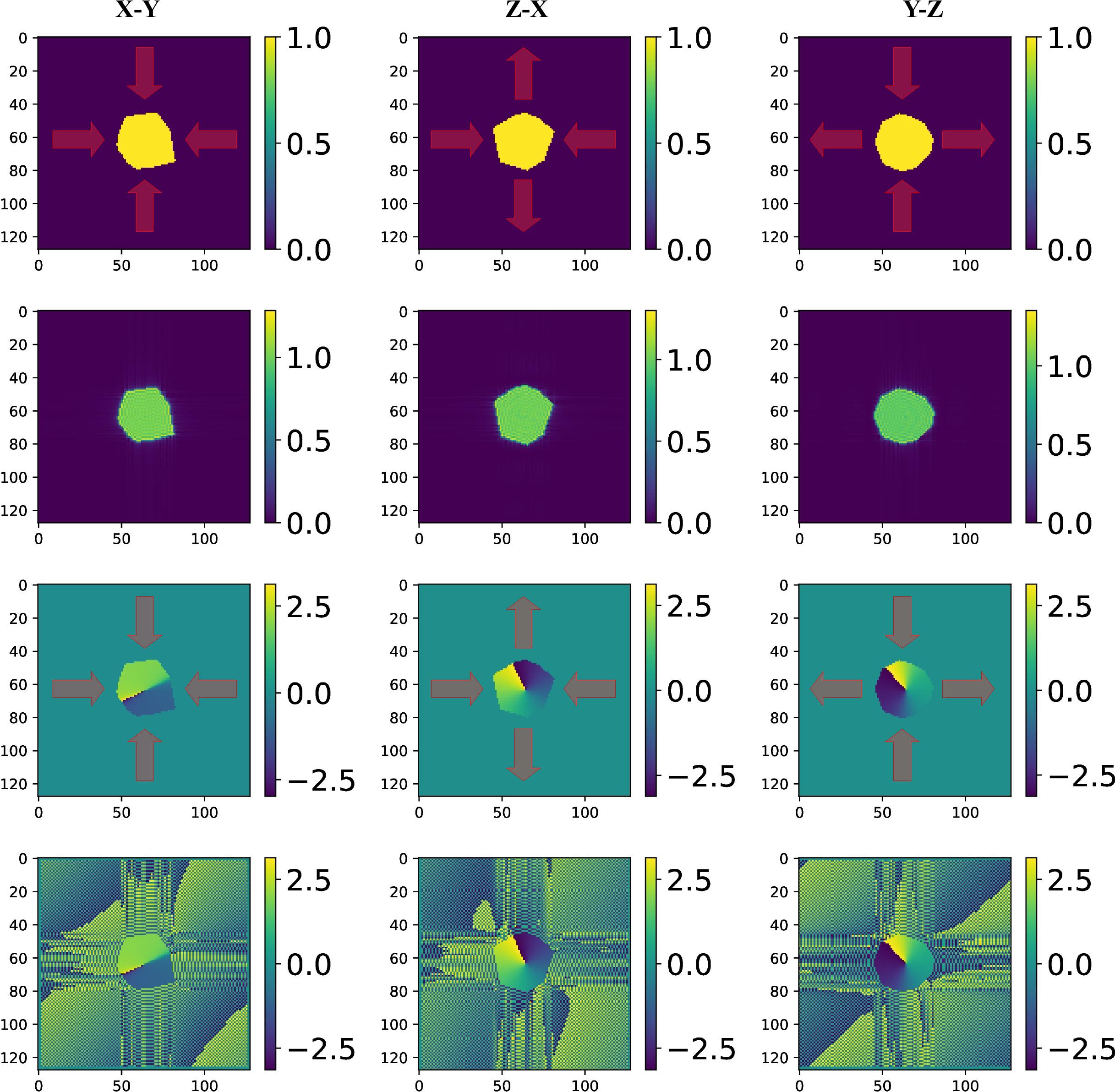}
	\caption{
		Fourier transform -based bulk resampling demonstrated on a synthetic crystalline volume with a screw dislocation, marked by a winding phase at the dislocation core. 
\textbf{Row 1}: orthogonal slices through center of the amplitude profile of the original crystal. The arrows denote the bulk compression/expansion along the principal axes with respect to the original object, as a result of resampling. 
		\textbf{Row 2}: central cross-sections of the resampled amplitude. 
\textbf{Row 3}: corresponding slices of the original phase, with the screw dislocation indicated by the phase discontinuity. \textbf{Row 4}: resampled phase profile. The plots show phase in radians.
}
	\label{fig.bulk_rescaling}
\end{figure}

Bulk resampling of a synthetic object originally sampled in steps of $s_0$ is shown in Fig.~\ref{fig.bulk_rescaling}. 
The first two rows show the amplitudes of the original and bulk-resampled object. 
The third and fourth rows show the corresponding complex phases. 
The winding phase profile is indicative of a screw dislocation passing through the center of the crystalline bulk (third row). 
We note the fidelity of the resampled phase to the original object, specifically the discontinuity at the dislocation core. 
This could not have been achieved with a simple interpolation within a voxel, as seen in present-day methods~\cite{Newton2020,Gao2021,Wilkin2021}. 
Further, there is little difference in the sizes of the original and resampled objects because of the proximity of each $B_{ii}$ to the chosen value of $s_0$. 
For demonstrative purposes, the fidelity of the bulk-resampling procedure to high-frequency phase features even with greatly exaggerated under/over-sampling steps is shown in the Appendix Section~\ref{A:exagg}.

We note the following: 
\begin{enumerate}
	\item	
		The concept of upsampling with the padding operator $\mathcal{P}$ has been explored in the context of image registration~\cite{GuizarSicairos2008}. 
	\item	
		The operator $\mathcal{B}$ in Eq.~\eqref{eq.bulkresampl} returns a new array of size $N_1\times N_2\times N_3$ pixels which contains the original object resampled along axis $i$ by a factor $B_{ii}/s_0$.
	\item	
		The upsampling enabled by Eq.~\eqref{eq.bulkresampl} when $B_{ii} < s_0$ does not actually result in better sampling of any sub-pixel features in the physical object. 
		The smallest feature dictated by the available Fourier content is merely sampled more finely. 
		This is owing to the fact that no new Fourier information was added in the process. 
		However, on truncation ($B_{ii} > s_0$), high-frequency information in Fourier space is truly discarded. 
		As expected, the downsampling operation results in a loss of spatial resolution.
\item	
While truncation and zero-padding are not differentiable operations by themselves, \emph{every pixel value} in the output array $\psi^\prime_{\bs{m}}$ is still related analytically to the pixels in $\psi_{\bs{m}}$ through the Fourier transform. 
		In the MR-BCDI problem, this analytical dependency extends back to the original unknowns, namely $A_{\bs{m}}$ and $\bs{u}_{\bs{m}}$ through multiplication and exponentiation, as well as each of the inferred diffraction patterns $\norm{\Psi^{(i)}_{\bs{m}}}^2$ and the loss function in Eq.~\eqref{eq.objfun_multi} itself. 
		Multiplication, exponentiation and the Fourier transform are all differentiable functions and have highly optimized implementations in modern auto-differentiation software.
		In the language of auto-differentiation, the computational graph is unbroken from $\mathcal{A}_{\bs{m}}$ and $\bs{u}_{\bs{m}}$ to $\mathcal{L}_\text{multi}$, even in the presence of a ``destructive" operation like truncation. 
Therefore this formulation lends itself easily to auto-differentiation, which has already started to underlie computational imaging applications with x-rays~\cite{Kandel2019,Scheinker2020,Cherukara2020,Harder2021}.
	\item
		The $\mathcal{B}$ operators in Eq.~\eqref{eq.bulkresmpl_full} commute with each other since the resampling along each of the three axes are independent operations.
\end{enumerate}

 \subsection{Shear resampling}				\label{SS:shear}		The bulk resampling method described in Section.~\ref{SS:bulk} was achieved with non-analytical operations such as zero-padding and truncation, which nevertheless connect the unknowns $A$ and $\bs{u}$ analytically to the inferred signal. 
In contrast, the shear transformations corresponding to the off-diagonal elements of $\bs{B}_\text{real}$ are truly differentiable in nature. 
As we show in this subsection, they can be achieved through a combination of 1-D FTs and phase ramps. 
The 2D shear of a general 3D function $\xi(\bs{x}) \equiv \xi(x, y, z)$ in the $yz$-plane is given by the following sequence of operations: 
\begin{align}
	\xi(x+\alpha y+\beta z, y, z) &=  \notag \\
		\underbrace{\int_\mathbb{R}dk~e^{\iota 2\pi kx}}_{\mathcal{F}_1^{-1}}
		&\underbrace{e^{\iota 2\pi k(\alpha y + \beta z)}}_\text{phase shift}
		\underbrace{\int_\mathbb{R}dx~e^{-\iota 2\pi kx}}_{\mathcal{F}_1}
		\xi(x, y, z)	\\
		&= \underbrace{\mathcal{F}_1^{-1} \Phi_{23}(k\left|\alpha, \beta\right.)\mathcal{F}_1}_\text{off-diagonal shear} \xi(x, y, z) \label{eq.shearoperator}
\end{align}
Here, $\Phi_{23}(k\left|\alpha, \beta\right.) \equiv \exp[\iota 2\pi k(\alpha y + \beta z)]$ is the phase ramp applied to axes 2 and 3 ( in this case, $y$ and $z$ respectively). 
Similar phase ramps $\Phi_{31}(k|\alpha, \beta)$ and $\Phi_{12}(k|\alpha, \beta)$ are constructed with the 1D Fourier transforms $\mathcal{F}_2$ and $\mathcal{F}_3$ to achieve shears along the $zx$- and $xy$-planes. 
These are given by: 
\begin{align}	
	\Phi_{31}(k | \alpha, \beta) &= \exp[\iota 2\pi (\alpha z + \beta x)] \\
	\Phi_{12}(k | \alpha, \beta) &= \exp[\iota 2\pi (\alpha x + \beta y)]
\end{align}

\begin{figure}\centering
	\includegraphics[width=0.65\textwidth]{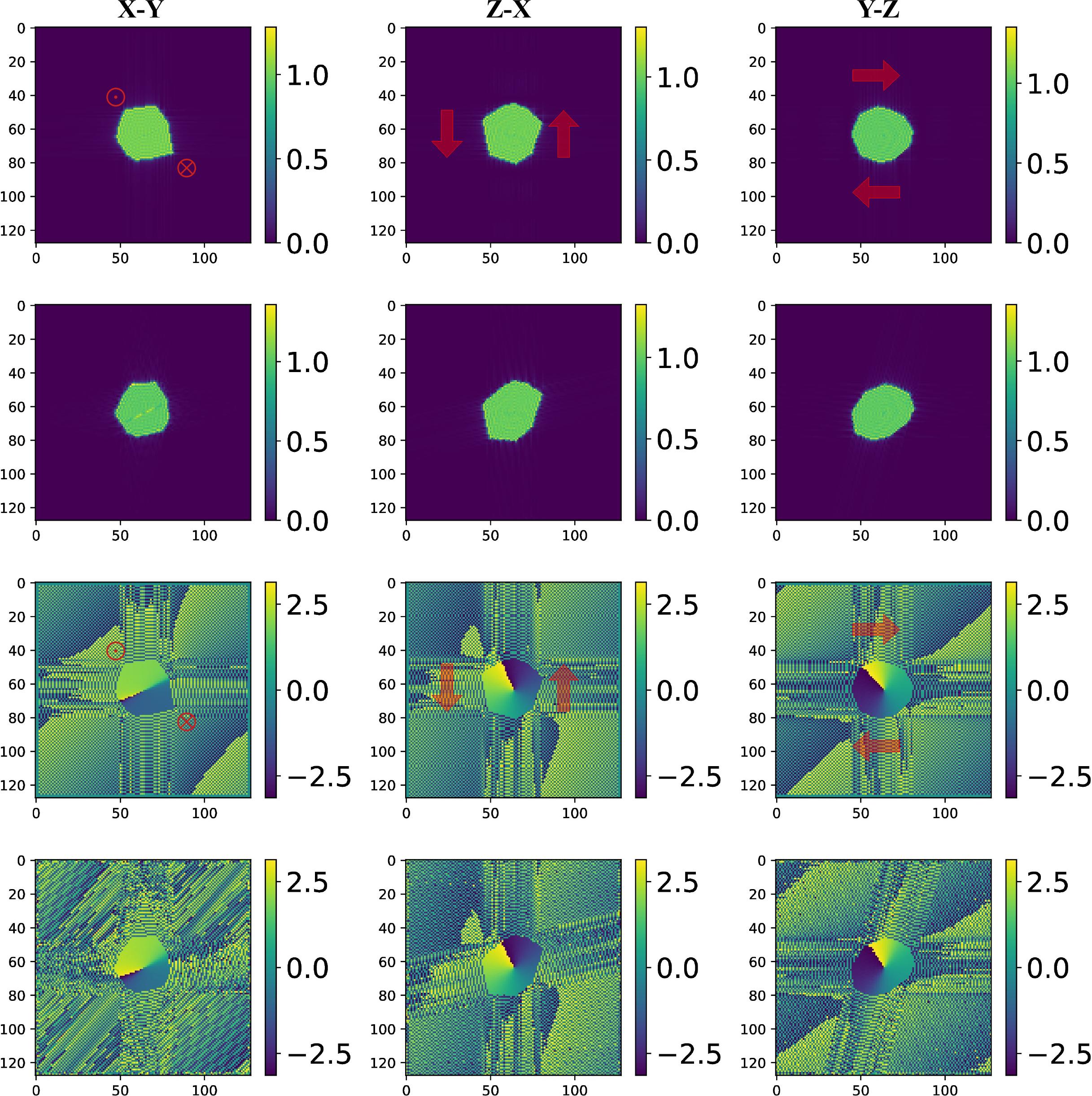}
	\caption{
		\textbf{Row 1}: amplitude cross-section of the \emph{bulk-resampled} digital object in Fig.~\ref{fig.bulk_rescaling}, before application of the shearing operator in Eq.~\eqref{eq.shearreasmpl_full}.
		The arrows denote the direction of shearing. 
		The $\otimes$ and $\odot$ symbols denote arrows entering and exiting the plane of the figure respectively. 
		\textbf{Row 2}: shear-resampled object amplitude cross-sections. 
		\textbf{Row 3,4}: corresponding phase profiles, indicating the screw dislocation. 
		All phase plots are in radians. 
	}
	\label{fig.shear_rescaling}
\end{figure}

The use of the sampling matrix $\bs{B}_\text{real}$ from Ref.~\cite{Maddali2020a} lends itself easily to the compact formulation of the shear operator in Eq.~\eqref{eq.shearoperator}. 
To see this, we consider the desired sampling step along $\unitvector{k}_1$, namely $B_{11}l + B_{12}m + B_{13}n$ (where $\bs{m} \equiv [l~m~n]^T \in \mathbb{Z}^3$). 
This may be obtained from $\psi^\prime_{\bs{m}}$ in  Eq.~\eqref{eq.bulkresmpl_full} by:
\begin{align}
	\psi^{\prime\prime}_{\bs{m}} \equiv \psi^\prime([B_{11}~&B_{22}~B_{33}]^T\bs{m}) \longrightarrow \notag \\
	&\psi^\prime([B_{11}l+B_{12}m+B_{13}n~B_{22}~B_{33}]^T\bs{m}) \notag \\
	&= \underbrace{\left[
			\mathcal{F}_1^{-1} \Phi_{23}\left(
				k \left| \frac{B_{12}}{B_{11}}, \frac{B_{13}}{B_{11}} \right.
			\right) \mathcal{F}_1
		\right]}_\text{$yz$ shear operator} \psi^\prime_{\bs{m}}
\end{align}
where the $yz$-shear operator acts not on the original objet $\psi(s_0\bs{m})$, but the \emph{bulk-resampled} object $\psi^\prime_{\bs{m}}$ from Eq.~\eqref{eq.bulkresmpl_full}. 
Consequently, the complete shear operation is given by the composition of the shear operators along the independent Cartesian directions: 
\begin{align}
	\psi^{\prime\prime}_{\bs{m}} &\equiv \psi(\bs{B}_\text{real}\bs{m}) \notag \\
	&=	\left[\shearresample{3}{1}{2}\right] \circ
		\left[\shearresample{2}{3}{1}\right] \circ
		\left[\shearresample{1}{2}{3}\right]
		\psi^\prime_{\bs{m}}
		\label{eq.shearreasmpl_full}
\end{align}

where $\circ$ denotes operator composition. 
Fig.~\ref{fig.shear_rescaling} shows the effect of the shear resampling operation in Eq.~\eqref{eq.shearreasmpl_full} applied to the bulk-resampled object (Fig.~\ref{fig.bulk_rescaling}, row 2). 
Once again we note the fidelity to the original discontinuity, which would have not been possible with a simple real-space interpolation within the space of a single voxel. 

Eq.~\eqref{eq.bulkresmpl_full} and Eq.~\eqref{eq.shearreasmpl_full} together denote the sequence of analytic operations that take the unknown object $\psi(s_0\bs{m})$ to the appropriately sampled object $\psi(\bs{B}_\text{real}\bs{m})$, for direct use in Eq.~\eqref{eq.objfun_multi}. 
The complete resampling operation $\psi(s_0\bs{m}) \longrightarrow \psi(\bs{B}_\text{real}\bs{m})$ \emph{differentiably} transforms the original object (Fig.~\ref{fig.bulk_rescaling}, row 1) to the appropriately resampled object (Fig.~\ref{fig.shear_rescaling}, row 2), while preserving the fidelity to high-frequency features within the object (in this case, phase discontinuities). 

\subsection{Resampling by grid rotation}	\label{SS:rotresampl}	Building upon the shear-resampling methodology in Section~\ref{SS:shear}, we describe our last fundamental resampling operation in which the discrete grid of the object is arbitrarily rotated. 
This is of crucial importance in the MR-BCDI problem, since the scatterer has to be rotated into different Bragg conditions in turn.
The method decomposes the desired rotation into a sequence of shear operations, and therefore has all the advantages of the shear-resampling method, including differentiability and maximum fidelity to small features.

As an example, we wish to actively rotate the discretized object $\psi(s_0\bs{m})$ by an angle $\theta$ about the axis $+\unitvector{k}_3$ (in a right-handed manner). 
Practically, we rotate the Cartesian sampling grid by $-\theta$ about $+\unitvector{k}_3$. 
According to Refs.~\cite{Unser1995,Larkin1997,Thevenaz2000}, Eq.~\eqref{eq.shearoperator} may be employed to achieve this by three successive shear operations: 
\begin{align}
	\mathcal{R}(\theta, \unitvector{k}_3)\psi(\bs{x}) &\equiv \psi\left(x\cos\theta+y\sin\theta, -x\sin\theta + y\cos\theta, z\right) \notag \\
		&=	\left[\mathcal{F}_1^{-1}\Phi_{12}\left(k\left|0, \tan\frac{\theta}{2}\right.\right)\mathcal{F}_1\right] \circ
			\left[\mathcal{F}_2^{-1}\Phi_{12}\left(k\left|-\sin\theta, 0 \right.\right)\mathcal{F}_2\right] \circ
			\left[\mathcal{F}_1^{-1}\Phi_{12}\left(k\left|0, \tan\frac{\theta}{2}\right.\right)\mathcal{F}_1\right] \psi(\bs{x})
			\label{eq.rotZ}
\end{align}
Similar formulations to Eq.~\eqref{eq.rotZ} hold for rotations $\mathcal{R}(\theta,\unitvector{k}_1)$ and $\mathcal{R}(\theta,\unitvector{k}_2)$ about the $\unitvector{k}_1$ and $\unitvector{k}_2$ directions. 
Using this method, we can express the grid resampling of an \emph{arbitrary} rotation $\mathcal{R}(\theta, \unitvector{n})$ as a composition of three rotations: 
\begin{equation}
	\mathcal{R}(\theta,\unitvector{n}) = 
		\mathcal{R}(\alpha_3,\unitvector{k}_3) \circ
		\mathcal{R}(\alpha_2,\unitvector{k}_2) \circ
		\mathcal{R}(\alpha_1,\unitvector{k}_1)
\end{equation}
Here the $\alpha_i$ denote the Euler angles of rotation in the $XYZ$ convention (\emph{i.e.}, rotation about $X$ followed by $Y$ and then $Z$). 
We note that regardless of the Euler convention, an arbitrary rotation may be achieved by a triplet of successive rotations about the principal Cartesian axes.

\begin{figure}\centering
	\includegraphics[width=0.65\textwidth]{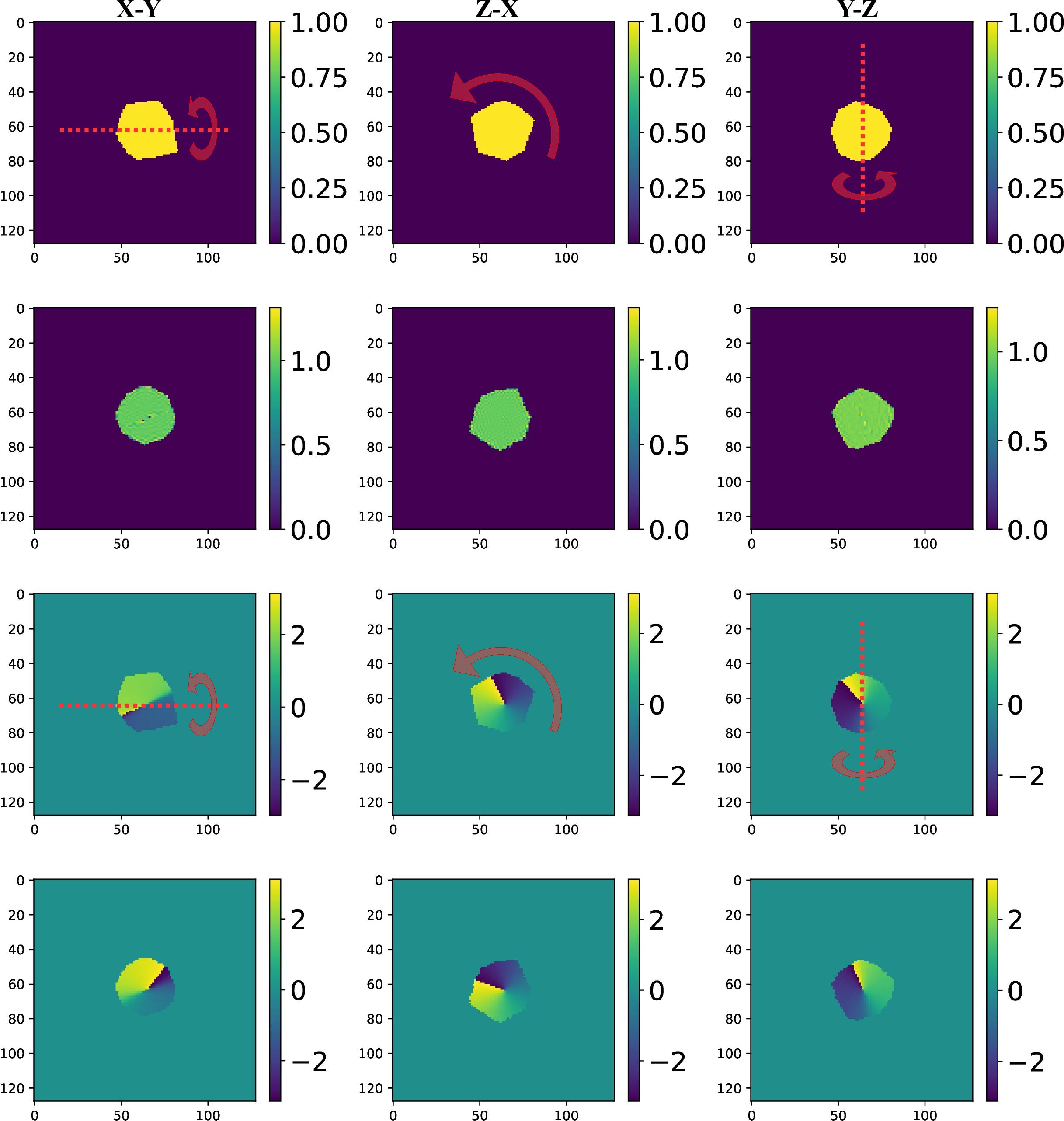}
	\caption{
		\textbf{Row 1}: Mutually orthogonal cross sections of the original scatterer amplitude. The arrows indicate the direction of the $45^\circ$ rotation about the $Y$-axis viewed from different perspectives. 
		\textbf{Row 2}: scatterer amplitude after it has been rotated within the grid. 
		\textbf{Row 3, 4}: corresponding complex phase cross sections. 
		All phase plots are in radians. 
	}
	\label{fig.rotresampl}
\end{figure}

The rotation of the example crystal from the previous examples by an angle of $45^\circ$ about the $Y$-axis is shown in Fig.~\ref{fig.rotresampl}. 
Again we note the fidelity the resampled object to the discontinuity in the bulk. 
The Appendix Section~\ref{A:rotate_resample} contains an example that contrasts Fourier-based rotation resampling with regular grid interpolation in a crystal containing multiple bulk discontinuities.

We note here in our global optimization scheme, $\mathcal{A}$ and $\bs{u}$ are defined and reconstructed only within an estimated bounding box $\mathcal{V}$ of the object. 
However, the transformations described in this section are applied to a zero-padded version of $\mathcal{V}$.
In this paper, all buffered arrays are of size $128\times 128\times 128$. 
While our interpolation approach ensures that phase discontinuities are well-preserved, we note that object amplitude suffers from Gibbs phenomenon -induced oscillations at sharp edges and discontinuities. 
This is evident on comparing the first and second rows in Figures~\ref{fig.bulk_rescaling} and~\ref{fig.shear_rescaling}. 
However, this problem is automatically addressed in our global optimization scheme through the use of the TV regularizer. 
As we have shown, this results in smoothed edges in the reconstructions. 

Further, we note that a large rotation angle about a principal axis (for example, $\theta$ about the $Z$-axis from Eq.~\eqref{eq.rotZ}), when implemented with the fast Fourier transform,  potentially causes the transformed object to be split across the periodic array boundary, due to an excessively large shear. 
Whether or not this splitting takes place depends on the size of $\theta$, and the ratio of the crystal-to-buffer sizes in each array dimension. 
A smaller ratio implies a larger $\theta$ can be accommodated before splitting commences. 
This undesirable effect can be addressed by performing a large principal rotation in a sequence of smaller rotations. 
In this work, we have conservatively set this smaller rotation size to $45^\circ$. 
With this convention, a rotation of $210^\circ$ about the $Z$-axis would be decomposed as four rotations of $45^\circ$ followed by a $30^\circ$ rotation.
While this certainly lengthens the chain of analytic computations for each optimization step, we have found this to introduce negligible overhead when implemented on the GPU. 
 
\appendix
\section{Mutual orthogonality}\label{A:MO}
Consider a $3 \times 3$ matrix $\bs{M} \equiv [\bs{v}_1~\bs{v}_2~\bs{v}_3]$ whose columns represent 3 linearly independent vectors spanning 3D Euclidean space $\mathbb{R}^3$. 
Here each of the $\bs{v}_i \in \mathbb{R}^3$ is a $3 \times 1$ column vector. 
The \emph{mutual orthogonality} (MO) of $\bs{M}$ is defined as: 
\begin{align}
    \mathcal{O}(\bs{M}) &\equiv \frac{\bs{v}_1 \cdot \bs{v}_2 \times \bs{v}_3}{\norm{\bs{v}_1}\norm{\bs{v}_2}\norm{\bs{v}_3}} \\
    &= \frac{\det(\bs{M})}{\norm{\bs{v}_1}\norm{\bs{v}_2}\norm{\bs{v}_3}}
    \label{eq.definition}
\end{align}
Here $\norm{\cdot}$ denotes the Euclidean distance or the $L_2$-norm. 
We see that $\mathcal{O}(\bs{M})$ is invariant under even permutations of the $\bs{v}_i$'s and changes sign but not magnitude under odd permutations. 
This follows simply from the same properties of the determinant $\det(\bs{M})$. 
A positive (negative) sign of $\mathcal{O}(\bs{M})$ indicates that the columns of $\bs{M}$ are right (left) -handed in order. 

If the basis vectors are mutually orthogonal (\emph{i.e.}, $\bs{v}_1 \perp \bs{v}_2 \perp \bs{v}_3$), then we see that $\mathcal{O}(\bs{M}) = \pm 1$, indicating a rectangular basis for $\mathbb{R}^3$. 
If they all lie in the same plane, $\mathcal{O}(\bs{M}) = 0$ and the `basis' does not in fact span $\mathbb{R}^3$. 
Thus, $\mathcal{O}(\bs{M}) \in [-1, 1]$ contains information about the relative independence of the basis vectors of $\bs{M}$, as well as their handedness. 

\begin{figure}
    \centering
    \includegraphics[width=0.3\textwidth]{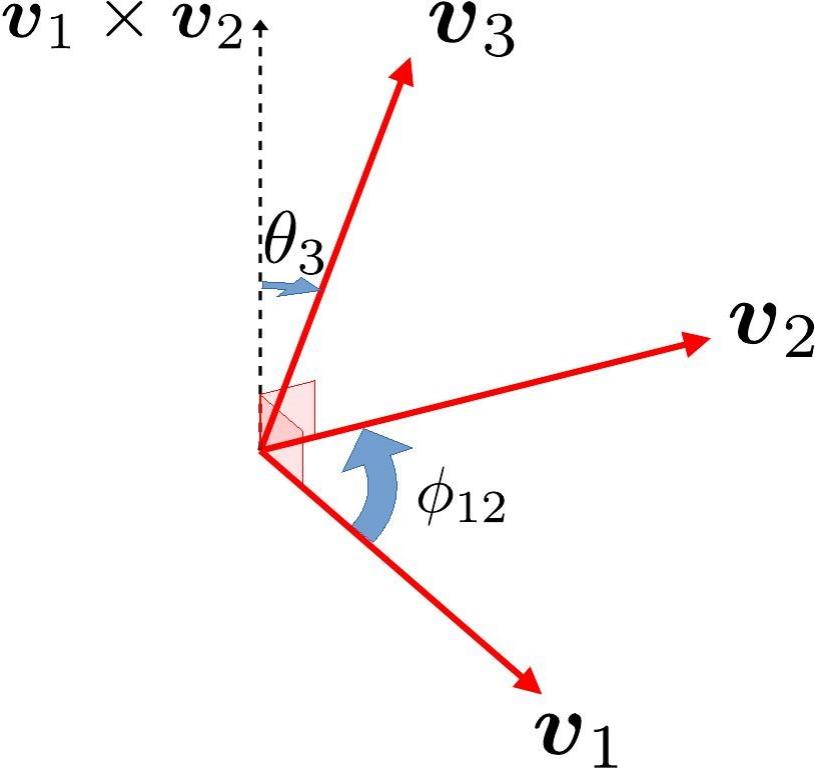}
    \caption{Schematic elucidating angle parameterization of MO for a basis matrix $\bs{M} = [\bs{v}_1~\bs{v}_2~\bs{v}_3]$. }
    \label{fig:parm}
\end{figure}

We refer to Fig.~\ref{fig:parm}.
Here, $\mathcal{O}(\bs{M})$ can be parameterized in terms of the angles between different vectors from the magnitudes of the cross and dot products: 
\begin{align}
    \mathcal{O}(\bs{M}) &= \frac{\norm{\bs{v}_3}\norm{\bs{v}_1\times\bs{v}_2}\cos\theta_3}
    {\norm{\bs{v}_1}\norm{\bs{v}_2}\norm{\bs{v}_3}} \\
    &= \frac{
        \norm{\bs{v}_3}
        \left(\norm{\bs{v}_1}\norm{\bs{v}_2}\sin\phi_{12}\right)
        \cos\theta_3}
    {\norm{\bs{v}_1}\norm{\bs{v}_2}\norm{\bs{v}_3}} \\
    &= \sin\phi_{12}\cos\theta_3 \label{eq.angparm}
\end{align}
We see that the MO of a basis of vectors does not depend on their magnitudes, but only their relative orientations. 
This parameterization is useful in determining the relation between the orthogonalities of mutually Fourier-conjugate bases, as we see below.

The formulation of MO as given above is particularly useful in BCDI.
The measurement consists of rectilinear (but not orthogonal) samples of x-ray intensity in reciprocal (Fourier) space, given by a $3 \times 3$ sampling basis matrix $\bq$~\cite{Maddali2020a,Li2020}. 
In other words, Fourier space is sampled at points $\bs{q}$ obtained from integer combinations of the basis vectors.
\begin{align}
    \bs{q} &= l\bs{v}_1 + m\bs{v}_2 + n\bs{v}_3 \\
	&= \bs{B}_\text{recip} \left[\begin{matrix}l\\m\\n\end{matrix}\right] \\
    \text{where } l, m, n &\in \mathbb{Z}. \notag
\end{align}
In practice, the integers $l, m, n$ range over $N_1, N_2, N_3$ values respectively: $l = 0, 1, \ldots, N_1-1$, and so on for $m$ and $n$ as well.  
Thus, the Fourier window is of size $N_1 \times N_2 \times N_3$ pixels. 

The columns of $\bq$ are determined by the diffraction geometry and the chosen manner of crystal rotation (`rocking'). 
This in turn unambiguously determines the conjugate basis $\br$ whose columns denote the real-space samplings steps of the object wave (\emph{i.e.}, the object itself). 
The real-space object sampled using $\br$ is related to the propagated wave sampled in the far field using $\bq$ through the discrete Fourier transform. 
The relation between the conjugate bases is given by~\cite{Maddali2020a}: 
\begin{align}
    \br = \bq^{-T} {\underbrace{\left[\begin{matrix}
    N_1 & & \\
     & N_2 & \\
     & & N_3
    \end{matrix}\right]}_{\mathcal{D} = \diag( N_1, N_2, N_3)}}^{-1}
    \label{eq.conj}
\end{align}
where we have denoted $\mathcal{D} \equiv \diag( N_1, N_2, N_3)$. 
Here, $(\cdot)^{-T}$ equivalently denotes the inverse of the transpose or the transpose of the inverse.
As we can see, a larger Fourier window (greater $N_1$, $N_2$, $N_3$) results in a better real-space resolution \emph{limit} (smaller sampling steps). 

The rotating crystal geometry in BCDI ensures that the diffraction signal is always sampled not in a well-behaved, rectangular manner, but with a linear shear. 
In other words, $\norm{\mathcal{O}(\bq)} < 1$. 
This naturally introduces a shear into the real-space sampling basis $\br$ through Eq.~\eqref{eq.conj}. 
In the light of this, we would like to determine the relation between $\mathcal{O}(\bq)$ and $\mathcal{O}(\br)$. 
Ideally, we would prefer $\mathcal{O}(\bq)$ for each scan in a multi-reflection BCDI measurement to be as close to unity as possible, and therefore one must arrange their measurements accordingly. 
In our simulations, we choose to exercise this flexibility rather coarsely through one of two rocking directions, namely $\theta$ and $\phi$ from Figure~\ref{fig.schematic}. 
These correspond to motor stages at the 34-ID-C coherent diffraction instrument at the Advanced Photon Source. 

We seek an expression for $\mathcal{O}(\br)$ in terms of $\mathcal{O}(\bq)$.  
In order to do this, we first define $\betaq \equiv \bq\mathcal{D}^{1/2}$ and $\betar \equiv \br\mathcal{D}^{1/2}$. 
This implies that Eq.~\eqref{eq.conj} can be rewritten more simply as: 
\begin{equation}
    \betar^T\betaq = \mathds{1}
    \label{eq.conj2}
\end{equation}
where $\mathds{1}$ is the $3 \times 3$ identity matrix. 
From the above definitions, we see that $\br$ ($\bq$) differs from $\betar$ ($\betaq$) only in the magnitudes of the columns, and therefore we have: $\mathcal{O}(\betaq) = \mathcal{O}(\bq)$ and $\mathcal{O}(\betar) = \mathcal{O}(\br)$. 
It is therefore sufficient to determine the relation between $\mathcal{O}(\betaq)$ and $\mathcal{O}(\betar)$.

If $\betaq \equiv [\bs{k}_1~\bs{k}_2~\bs{k}_3]$ and $\betar \equiv [\bs{r}_1~\bs{r}_2~\bs{r}_3]$, then Eq.~\eqref{eq.conj2} tells us that: 
\begin{align}
    \bs{r}_l &= \frac{\epsilon_{lmn} \bs{k}_m \times \bs{k}_n}{\det(\betaq)} 
    = \frac{\epsilon_{lmn}\bs{k}_m \times \bs{k}_n}
    {\bs{k}_1 \cdot \bs{k}_2 \times \bs{k}_3}
    ~\text{for } l, m, n = 1, 2, 3 \\
    \Longrightarrow
    \norm{\bs{r}_l} &= \frac{1}{\norm{\bs{k}_l} \cos\theta_l} \tag{from the geometry in Fig.~\ref{fig:parm}}
\end{align}
Here $\epsilon_{lmn}$ is the Levi-Civita symbol denoting the cyclic relation between the indices $l$, $m$ and $n$. 
We therefore have: 
\begin{align}
    \mathcal{O}(\betar) &= \frac{\det\left(\betar\right)}{\norm{\bs{r}_1}\norm{\bs{r}_2}\norm{\bs{r}_3}} \\
    &= \frac{\norm{\bs{k}_1}\norm{\bs{k}_2}\norm{\bs{k}_3}}{\det\left(\betaq\right)}
    \cos\theta_1\cos\theta_2\cos\theta_3 \tag{$\because\det\betar = 1/(\det\betaq)$ from Eq.~\eqref{eq.conj2}} \\
    \Longrightarrow &\boxed{
        \mathcal{O}(\betar) = \frac{\cos\theta_1\cos\theta_2\cos\theta_3}{\mathcal{O}(\betaq)} \label{eq.finalresult}
    }
\end{align}

We note that a high MO in reciprocal space indicates a similarly high MO in real space. 
These MOs respectively pertain to how the signal and the scatterer are discretized. 
Hence choosing  $\mathcal{O}(\bq)$ as close to $1$  as possible by judicious choice of crystal rotation (in particular, rocking about the $\theta$ or $\phi$ axis) facilitates the best orthogonal discretization of the reconstructed object wave, owing to interpolation from a minimally sheared sampling basis $\bs{B}_\text{real}$. 
 
\section{Silicon carbide nanoparticle fabrication}\label{A:SiC}
The SiC nanoparticles were fabricated from a single 4H-SiC wafer with an \emph{i-p-n} structure procured from Norstel. 
The wafer comprised a $400$ nm-thick intrinsic SiC layer with $<10^{15}$ cm$^{-3}$ intrinsic defect content, a $2~\mu$m-thick \emph{p}-type SiC layer with $<10^{19}$ cm$^{-3}$ of aluminum atoms, and a $0.5$ mm-thick \emph{n}-type handle with $<10^{18}$ cm$^{-3}$ of nitrogen atoms. 
The wafer was diced into $5\times5$ mm$^2$ chips, patterned by electron beam lithography (bi-layer PMMA 495K-A3/950K-A6; $1000\mu$C cm$^{-2}$ dose, 3-minute development in 1:3 MIBK:IPA at room temperature), and metallized (10 nm of Ti and 60 nm of Cr via e-beam evaporation) to transfer the asymmetric, D-shaped arrays onto the SiC using plasma etching. 
During synthesis, the D shape flat was aligned to the crystallographic $[11\bar{2}0]$ direction (see Figure~\ref{fig.sic_iso}(d)). 
The SiC was etched (SF6-Ar ICP/RIE dry-etching) with the Ti/Cr acting as a hard mask. 
The $1.2~\mu$m etch depth exposed the \emph{p}-type layer for subsequent photo-electrochemical (PEC) etching of the nanoparticles. 
The plasma etch-defined nanopillars were then PEC etched (0.2M KOH, -0.3V biased etch under 365nm UV illumination at  $<500$ mW) to partially and selectively remove the \emph{p}-type SiC to allow for nanoparticle formation and subsequent detachment. 
For this particular experimental run, the \emph{p}-type SiC was underetched, resulting in a more substantial, weakly scattering, porous \emph{p}-type tether remaining. 
This resulted in the nanoparticle having a larger perceived size as seen in Figure~\ref{fig.sic_iso}(e).
These partially undercut nanopillar/nanoparticle arrays were stamp-transferred to a PMMA-coated silicon substrate. 
The wafers were baked at the PMMA glass transition temperature, which locked the nanoparticles on the Si substrate. 
The trasnferred nanoparticles were arranged on the Si wafer in $100~\mu$m pitched arrays bordered by macroscopic fiducials consisting of tightly packed nanoparticles. 
Finally, the PMMA was removed by oxygen plasma, leaving pristine SiC on Si nanoparticle arrays that were then conformally covered with alumina (22 nm thick) via ALD.

\section{Phase retrieval recipe}\label{A:recipe}
Given below is the phase retrieval recipe used in all the reconstructions in this paper. 
All these algorithms are reviewed in Ref.~\cite{Marchesini2007}. 
\begin{enumerate}
	\item	$150$ iterations of ER, with shrinkwrap every $30$ iterations. 
	\item	$300$ iterations of HIO. 
	\item	$100$ iterations of solvent-flipping (SF), with shrinkwrap every $25$ iterations. 
	\item	$300$ iterations of HIO. 
	\item	$450$ iterations of ER, with shrinkwrap every $90$ iterations. 
\end{enumerate}
 
\section{Feature fidelity with exaggerated resmpling}\label{A:exagg}
\begin{figure}[!ht]
	\centering
	\includegraphics[width=0.75\textwidth]{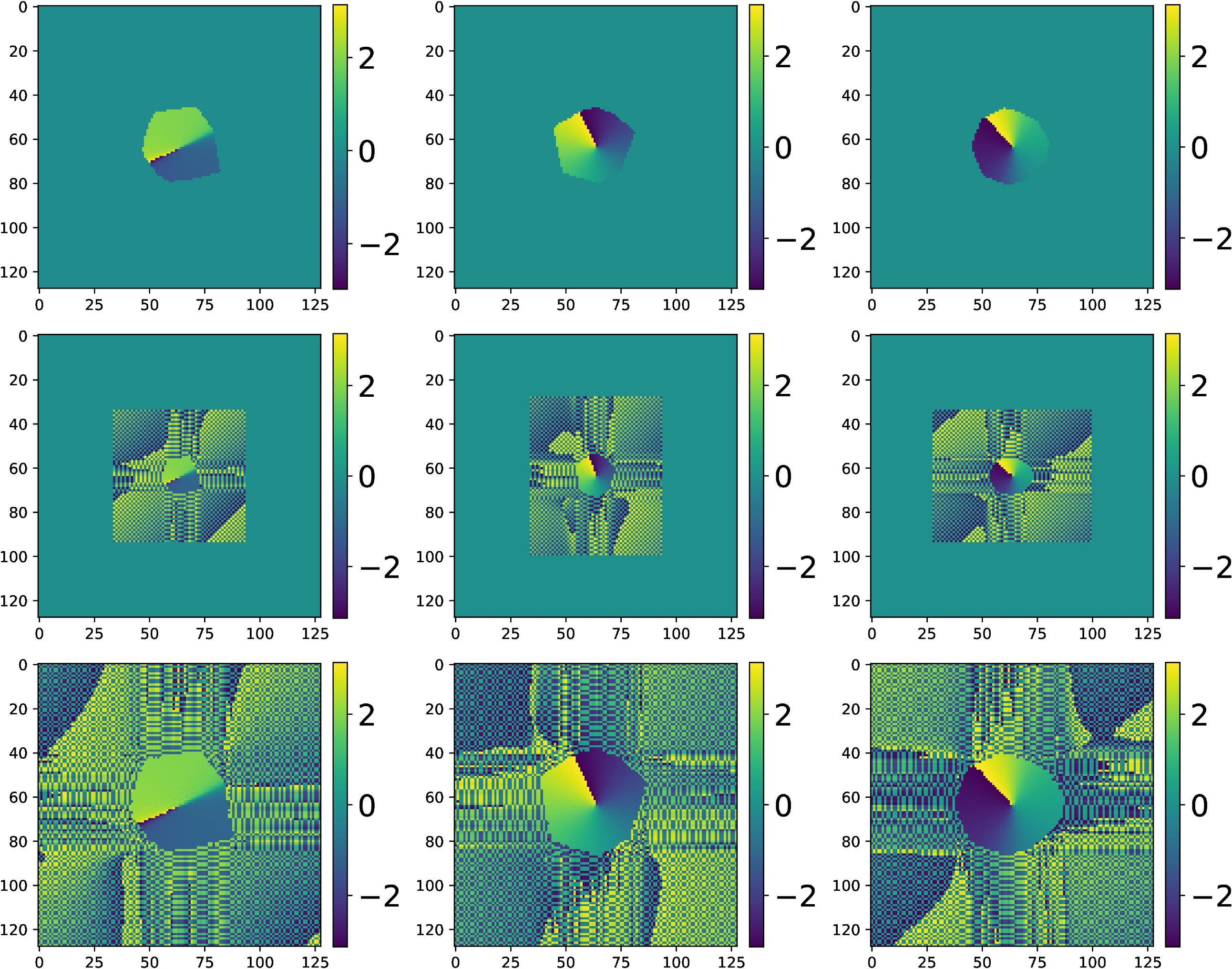}
	\caption{
		Orthogonal cross sections of a discontinuous 3D phase object subjected to uniform  bulk transformation $\bulkresample{i}{\xi}$. 
		For simplicity, $s_0 = B_{ii} = 1$ for all $i = 1, 2, 3$ (along all three axes),  
		with (\textbf{top row}) $\xi = 1$ (original), \textbf{(middle row)} $\xi = 1.5$ and (\textbf{bottom row}) $\xi = 0.75$. 
	}
	\label{fig.bulk_exag}
\end{figure}

Visual demonstration of feature fidelity with exaggerated under- and over-sampling (Figure~\ref{fig.bulk_exag}). 
 
\section{Sequential optimization plans}\label{A:optimplans}
\begin{table}[!ht]
\centering
\caption{
	Sequential optimization plan for dislocation-free crystal. 
	The BCDI scans within each minibatch were randomly selected. 
}    
\label{tab:optimplan1}
   \begin{tabular}{|c|c|c|c|}
   \hline
	\textbf{Epoch} & \textbf{Number of minibatches} & \textbf{Scans in minibatch} & \textbf{Iterations per minibatch} \\ \hline
	1 & 400	& 2	& 6		\\	\hline
	2 & 320	& 3	& 12	\\	\hline
	3 & 240	& 3	& 25	\\	\hline
	4 & 160	& 3	& 50	\\	\hline
	5 & 80	& 3	& 100	\\	\hline
	6 & 40	& 4	& 200	\\	\hline
	7 & 20	& 4	& 400	\\	\hline
	8 & 1	& 5 \textbf{(full dataset)}	& 2000	\\	\hline
   \end{tabular}
\end{table}

\begin{table}[!ht]
\centering
\caption{
	Optimization plan for crystal with orthogonal screw dislocations along $[111]$ and $[2\bar{2}0]$. 
	The BCDI scans within each minibatch were randomly selected. 
}    
\label{tab:optimplan2}
   \begin{tabular}{|c|c|c|c|}
   \hline
	\textbf{Epoch} & \textbf{Number of minibatches} & \textbf{Scans in minibatch} & \textbf{Iterations per minibatch} \\ \hline
	1 & 800	& 2	& 3		\\	\hline
	2 & 400	& 2	& 6		\\	\hline
	3 & 320	& 2	& 12	\\	\hline
	4 & 240	& 3	& 25	\\	\hline
	5 & 160	& 3	& 50	\\	\hline
	6 & 80	& 3	& 100	\\	\hline
	7 & 40	& 3	& 200	\\	\hline
	8 & 20	& 3	& 400	\\	\hline
	9 & 1	& 4 \textbf{(full dataset)}	& 2000	\\	\hline
   \end{tabular}
\end{table}

\begin{table}[!ht]
\centering
\caption{
	Optimization plan for the silicon carbide nanocrystal. 
	The BCDI scans within each minibatch were randomly selected. 
}    
\label{tab:optimplan3}
   \begin{tabular}{|c|c|c|c|}
   \hline
	\textbf{Epoch} & \textbf{Number of minibatches} & \textbf{Scans in minibatch} & \textbf{Iterations per minibatch} \\ \hline
	1 & 800	& 4	& 6		\\	\hline
	2 & 640	& 4	& 12		\\	\hline
	3 & 480	& 4	& 25	\\	\hline
	4 & 320	& 5	& 50	\\	\hline
	5 & 160	& 5	& 100	\\	\hline
	6 & 80	& 5	& 200	\\	\hline
	7 & 40	& 5	& 400	\\	\hline
	8 & 1	& 6 \textbf{(full dataset)}	& 2000	\\	\hline
   \end{tabular}
\end{table}
 
\section{Comparison of interpolation methods}\label{A:rotate_resample}
\begin{figure}[!ht]
	\centering
	\includegraphics[width=0.9\textwidth]{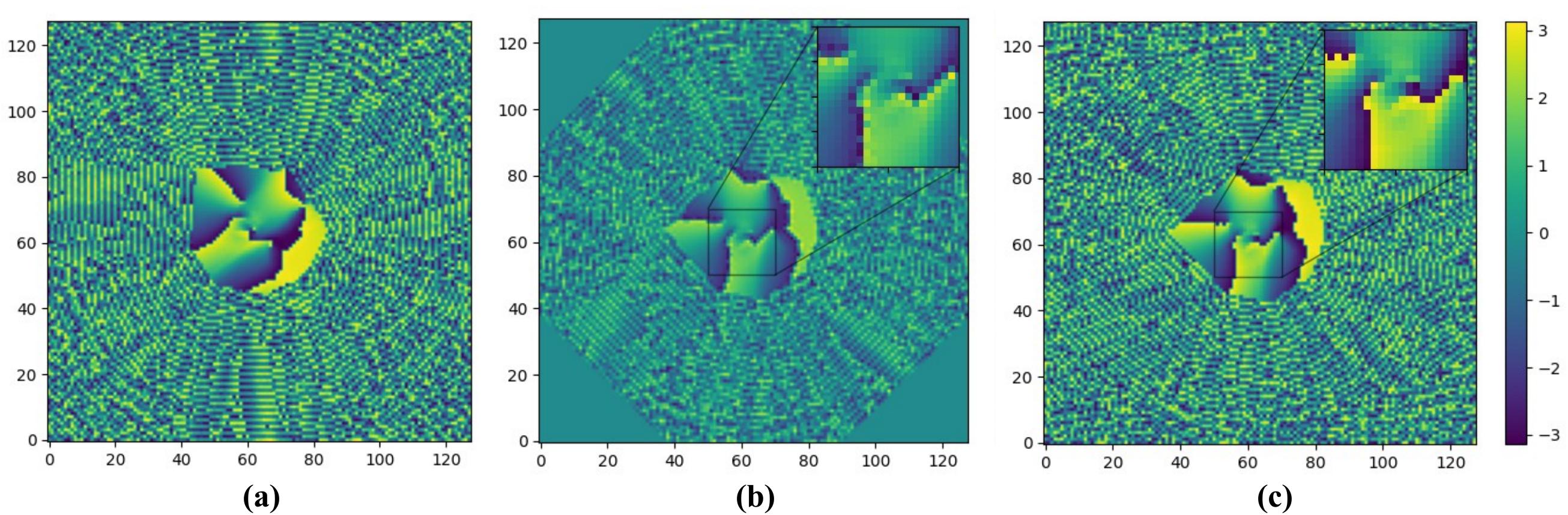}
	\caption{
		\textbf{(a)} 
			2D slice of 3D discontinuous  phase object prior to resampling by grid rotation. 
		\textbf{(b)} 
			Grid rotation by simple multi-linear interpolation between neighboring voxels. 
		\textbf{(c)} 
			Grid rotation by Fourier transform -based interpolation (Eq.~\eqref{eq.rotZ}). 
			The inset plots highlight the difference in fidelity to small, pixel-scale features. 
	}
	\label{fig.fouriervsinterp}
\end{figure}
 
\section{Funding} 
The development of the analytical MR-BCDI model and the Fourier transform-based interpolation method, proof-of-concept demonstration and design and fabrication of the SiC nanoparticles was supported by the US Department of Energy (DOE), Office of Science, Basic Energy Sciences, Materials Science and Engineering Division. 
Additional support for materials preparation came from the Q-NEXT Quantum Center, a U.S. Department of Energy, Office of Science, National Quantum Information Science Research Center, under Award Number DE-FOA-0002253.  
Silicon carbide deterministic nanoparticle fabrication and SEM characterization work was performed under proposals 72483 and 775514 in the Center for Nanoscale Materials clean room. 
Work performed at the Center for Nanoscale Materials, a U.S. Department of Energy Office of Science User Facility, was supported by the U.S. DOE, Office of Basic Energy Sciences, under Contract No. DE-AC02-06CH11357.
Refinement of the geometric, computational and optimization concepts was supported by the European Research Council (European Union’s Horizon H2020 research and innovation program grant agreement No. 724881). 
Generation of the simulated structures and the BCDI data acquisition was supported by the Laboratory Directed Research and Development (LDRD) funding from Argonne National Laboratory, provided by the Director, Office of Science, of the U.S. Department of Energy under Contract No. DE-AC02-06CH11357.
This research uses the resources of the Advanced Photon Source, a US DOE Office of Science User Facility operated for the DOE Office of Science by Argonne National Laboratory under contract No. DE-AC02-06CH11357.
 
\section{Acknowledgments} 
The authors gratefully acknowledge numerous valuable discussions with Drs. Anthony Rollett, Robert Suter and Matthew Wilkin (Carnegie Mellon University), Nicholas Porter and Dr. Richard Sandberg (Brigham Young University), Dr. Ross Harder (Argonne National Laboratory) and Dr. Anastasios Pateras (DESY). 
 
\section{Disclosures} 
The authors declare no conflicts of interest. 
Data and analysis scripts beyond those available in the code repository will be made available upon reasonable request.

\end{document}